\documentclass[fleqn,usenatbib]{mnras}

\usepackage{newtxtext,newtxmath}
\usepackage{subcaption}
\usepackage[T1]{fontenc}

\DeclareRobustCommand{\VAN}[3]{#2}
\let\VANthebibliography\thebibliography
\def\thebibliography{\DeclareRobustCommand{\VAN}[3]{##3}\VANthebibliography}

\usepackage{graphicx}	% Including figure files
\usepackage{amsmath}	% Advanced maths commands

\title[Excessive Bars in TNG50 ETGs]{Revisiting the Excess of Bar-like Structures in TNG50 Early-type Galaxies: Consistency and Tension with Observations}

\author[Hangci Du et al.]{
Hangci Du,$^{1, 2, 3}$\thanks{hcdu@bao.ac.cn}
Yougang Wang,$^{1,2}$\thanks{wangyg@bao.ac.cn}
Junqiang Ge$^{1,2}$\thanks{jqge@nao.cas.cn}
\\
$^{1}$National Astronomical Observatories, Chinese Academy of Sciences, Beijing 100101, China\\
$^{2}$School of Astronomy and Space Science, University of Chinese Academy of Sciences, Beijing 100049, China\\
$^{3}$Innovation Academy for Microsatellites, Chinese Academy of Sciences, Shanghai 201304, China
}

\date{Accepted XXX. Received YYY; in original form ZZZ}

\pubyear{\the\year{}}

\begin{document}
\label{firstpage}
\pagerange{\pageref{firstpage}--\pageref{lastpage}}
\maketitle

\begin{abstract}
The IllustrisTNG simulation suite, particularly TNG50, was reported to have generated a notable population of elongated, bar-like structures within galaxies classified as Early-Type Galaxies (ETGs). In this work, we revisit the nature of these structures at $z=0$ using a morphology-agnostic census. We find that these features are ubiquitous ($f_{\rm bar} \sim 75-80\%$) in dispersion-dominated galaxies ($D/T < 0.2$) in TNG50-1. They are not prolate rotators (rotating around their long axis), but genuine non-axisymmetric instabilities characterized by coherent, albeit slow, pattern speeds. Unlike the fast bars found in Late-Type Galaxies, these bar-like structures in ETGs are physically longer ($\gtrsim 3$ kpc), rotate significantly slower ($\Omega_p \lesssim 20$ km s$^{-1}$ kpc$^{-1}$), and reside in red, gas-poor, dispersion-dominated systems. By tracing the evolutionary history of these systems, we demonstrate that such structures originate as typical fast bars in gas-richer discs at higher redshifts ($z \gtrsim 0.2$). They survive the galaxy quenching phase, undergoing secular deceleration and lengthening due to dynamical friction, ultimately appearing as slow, fossilized rotators in the $z=0$ red sequence. We conclude that the specific excess of bar-like structures in TNG50 ETGs likely reflects a combination of the imperfect baryonic physics of the simulation (over-producing these bar-like structures or their host ETGs) and a potential observational blind spot regarding long-lived, secularly evolved bars in hot stellar systems.
\end{abstract}


%%%%%%%%%%%%%%%%%%%%%%%%%%%%%%%%%%%%%%%%%%%%%%%%%%

%%%%%%%%%%%%%%%%% BODY OF PAPER %%%%%%%%%%%%%%%%%%

\section{Introduction} \label{sec:intro}

Stellar bars represent one of the most ubiquitous non-axisymmetric structures in the Universe and serve as fundamental drivers of galactic secular evolution. By redistributing angular momentum and driving gas inflows \citep{2020MNRAS.499.1406L}, bars enhance central star formation, fuel supermassive black holes, and may eventually facilitate galaxy quenching \citep{2020MNRAS.492.4697N, 2020MNRAS.499.1116F, 2021A&A...651A.107G}. Consequently, they play a pivotal role in reshaping the structural and chemical abundance distributions of galaxies \citep{2019MNRAS.488L...6F}.

There remains an ongoing debate within the community regarding the preferential properties of bar-hosting galaxies. Specifically, it is contested whether bars are more frequently associated with red, rotation-supported systems or with blue, dispersion-dominated hosts.

Historically, the presence of bars has been inextricably linked to rotationally supported disc galaxies. Theoretical frameworks typically describe bar formation as a global instability within dynamically ``cold", rotation-dominated stellar discs \citep{1964ApJ...139.1217T}, a process potentially inhibited by a significant central spheroidal component \citep{1973ApJ...186..467O}. Accordingly, observational censuses have almost exclusively relied on morphological pre-selection, restricting samples to disc galaxies (e.g. via S\'ersic index cuts or axis ratio criteria) and implicitly treating dispersion-dominated systems as naturally bar-free \citep[e.g.][]{2007ApJ...659.1176M, 2009A&A...495..491A, 2012MNRAS.424.2180M}. Consistent with this view, early studies reported a higher bar fraction in low-mass, blue disc galaxies \citep{2008ApJ...675.1194B, 2009A&A...495..491A}.

However, subsequent observations have challenged this picture, indicating that bars are more prevalent in more massive, red, and dispersion-supported galaxies \citep{2012ApJ...750..141L, 2012MNRAS.424.2180M, 2012MNRAS.423.1485S, 2016A&A...595A..67C}. Furthermore, bar presence shows a significant anti-correlation with specific star formation rate and atomic hydrogen (H{\sc i}) gas fraction \citep{2013ApJ...779..162C, 2012MNRAS.424.2180M, 2017ApJ...835...80C}. Attempts to reconcile these conflicting trends suggest that the bar distribution may be bimodal \citep{2010ApJ...714L.260N, 2016A&A...587A.160D, 2025MNRAS.542..151M}, with a notable deficit of bars in intermediate-type or ``green valley'' galaxies \citep{2021JCAP...06..045D}.

Cosmological hydrodynamic simulations provide an invaluable laboratory for addressing these observational discrepancies, tracing the coupled evolution of discs, bulges, gas accretion, mergers, and feedback within a self-consistent $\Lambda$CDM framework. Nevertheless, early cosmological simulations frequently failed to reproduce the observed population of stellar bars—a challenge termed the ``bar problem" \citep[e.g.][]{2017MNRAS.469.1054A, 2019MNRAS.483.2721P}. This deficit was typically attributed to excessive gravitational softening scales \citep{2019MNRAS.483.2721P} and overly strong AGN feedback \citep{2014MNRAS.445..175G}. 

IllustrisTNG suite features optimized resolution and feedback mechanisms, enabling the robust capture of non-axisymmetric structures within a full cosmological context \citep{2018MNRAS.474.3976G, 2019MNRAS.490.3196P, 2019MNRAS.490.3234N, 2020MNRAS.491.2547R}. These simulations have achieved remarkable success in reproducing observed galactic morphologies \citep[e.g.][]{2019MNRAS.483.4140R, 2024MNRAS.535.1721P}. 

Across IllustrisTNG, bars preferentially form in galaxies with early ($z \sim 0.5$--$1.5$) disc assembly and older stellar populations \citep{2020MNRAS.491.2547R, 2022MNRAS.512.5339R, 2025A&A...697A.236L}. Gas  reservoir depletion is a critical prerequisite, with a threshold of $f_{\rm gas} \lesssim 0.4$ consistently identified as necessary for triggering instability \citep{2020ApJ...895...92Z, 2021A&A...647A.143L}. Once formed, bars act as engines of secular evolution, facilitating nuclear quenching via gas inflow and AGN fuelling \citep{2020MNRAS.491.2547R, 2024MNRAS.527.3366K}, while their pattern speeds decrease in accordance with angular momentum exchange theory \citep{2024A&A...692A.159S}. 

However, tensions still remain. Bars in TNG50 are systematically shorter than those in surveys like ATLAS$^{\rm 3D}$ or MaNGA \citep{2020ApJ...904..170Z, 2022ApJ...940...61F}, which is partly attributed to wavelength-dependence of bar measurements \citep{2025MNRAS.542.3154G}. We cannot rule out the imperfect-feedback effect \citep{2021A&A...650L..16F} and other unknown mechanisms neither.

Another tension between IllustrisTNG and observations, which remains underappreciated, is the prevalence of elongated, bar-like structures \citep{2020A&A...641A..60P, 2021A&A...647A.143L} in Early-Type Galaxies (ETGs). 

This population of ETGs in TNG50 is originally reported by \citet{2020A&A...641A..60P}. After their ETG pre-selection using cuts in the color-mass diagram, this odd populations resides in a ``forbidden" region of the $\lambda_{R} - \varepsilon$ diagram \citep[to the right of the magenta line defined by][]{2007MNRAS.379..418C}. These galaxies exhibit high ellipticity despite low rotational support, appear redder than the average population at fixed mass, and possess centrally elongated morphologies distinct from classical discs. This feature is not observed (as shown in  \citealt{2020A&A...641A..60P} section~5.2) in previous hydrodynamical simulations such as EAGLE \citep{2020MNRAS.494.5652W}, Magneticum \citep{2018MNRAS.480.4636S} and Illustris. To reconcile the simulation with observations, \citet{2020A&A...641A..60P} opted to excise these objects from their analysis by applying an axis ratio cut of $b/a < 0.6$. While this removes the tension, simply discarding a subset of the simulated population without a clear physical justification risks masking potential insights into the galaxy formation model. 

The physical nature of these objects remains debated. \citet{2020A&A...641A..60P} hypothesized that they may represent physically realistic systems that are ``failed discs"—galaxies that began to form a disc but whose evolution was disrupted by violent dynamics or intense feedback. Similarly, \citet{2021A&A...647A.143L} identified a population of ``bar-like galaxies" in TNG50 characterized by a rotating bar without an accompanying extended disc. They distinguished these from the classical ``bars embedded in discs" studied in TNG by other authors \citep[e.g.][]{2020MNRAS.491.2547R, 2020ApJ...895...92Z, 2020ApJ...904..170Z}.

Consequently, we adopt a morphology-agnostic strategy, diverging from notable prior studies of barred galaxies in TNG50 \citep[e.g.][]{2020MNRAS.491.2547R, 2020ApJ...895...92Z, 2020ApJ...904..170Z, 2022MNRAS.512.5339R}. Those studies typically applied pre-filters to mimic observational constraints, effectively assuming \textit{a priori} that barred galaxies must be disc galaxies. Furthermore, unlike \citet{2020A&A...641A..60P} and \citet{2021A&A...647A.143L}, who limited their samples to ETGs or specific ``bar-like galaxies'', we avoid artificially segregating the central elongated structures based on host morphology. Instead, we treat them as a continuous population across all morphological types. We refer to these features broadly as ``bar-like structures" to encompass both classical bars in discs and the elongated, rotating structures in quenched systems. By analyzing the full continuous population, we would investigate whether the distinct division between ``bar-like galaxies" in ETGs and ``barred spirals" in LTGs emerges naturally with no \textit{a priori} in TNG50, or if they are in fact continuously connected within the parameter space. More broadly, we aim to disentangle the factors contributing to the tension between TNG50 and observations regarding bar-like structures in ETGs. This involves determining whether the discrepancy stems from intrinsic limitations within the simulations or, conversely, whether bar-like structures are actually more frequent in observed ETGs than previously reported.

This paper is organized as follows. In Section \ref{sec:methods}, we detail the TNG50 simulation, our sample selection, and the methodology used to identify and characterize bar-like structures in a morphology-agnostic manner. In Section \ref{sec:results}, we present the statistical demographics of bar-like structures, comparing their abundance and properties against observational benchmarks, and offering a detailed anatomical comparison between canonical bars in discs and the disputed structures in dispersion-dominated systems. Section \ref{sec:discussion} discusses the physical drivers of these features, focusing on the link between angular momentum and stability, and addresses whether these structures represent a numerical artifact or an observational blind spot. Finally, we summarize our conclusions in Section \ref{sec:conclusions}.

\section{Methods} \label{sec:methods}

\subsection{Sample from TNG50}

Our investigation is conducted within the framework of the TNG50-1 simulation, the highest-resolution realization of the IllustrisTNG project \citep{2019MNRAS.490.3234N, 2019MNRAS.490.3196P}. TNG50 is a cosmological gravo-magnetohydrodynamical simulation performed with the moving-mesh code \texttt{AREPO} \citep{2010MNRAS.401..791S}. It is designed to trace the formation and evolution of galaxies from $z=127$ to the present day in a cubic comoving volume of $(51.7 \, \text{Mpc})^3$. The simulation assumes a $\Lambda$CDM cosmology consistent with the Planck 2015 results (Planck Collaboration et al. 2016), with parameters $\Omega_m = 0.3089$, $\Omega_b = 0.0486$, $\Omega_{\Lambda} = 0.6911$, $\sigma_8 = 0.8159$, $n_s = 0.9667$, and $h = 0.6774$. TNG50-1 contains approximately 20,000 resolved galaxies with stellar masses $M_* \gtrsim 10^7 M_{\odot}$, spanning a wide range of environments and morphologies. The simulation achieves a baryonic (gas and stellar) mass resolution of $8.5 \times 10^4 M_{\odot}$ and a dark matter mass resolution of $4.5 \times 10^5 M_{\odot}$ \citep{2024MNRAS.535.1721P}. This resolution is complemented by a high spatial fidelity; the gravitational softening length for stellar and dark matter particles is fixed to 288\,pc (physical) at $z \le 1$, while the adaptive softening for gas cells can reach a minimum of 72\,pc \citep{2019MNRAS.490.3196P}. This level of detail is crucial for resolving the internal dynamics and non-axisymmetric structures, such as bars, that are the focus of this work.

Our parent sample consists of all 903 galaxies at $z=0$ with stellar mass $M_* \ge 10^{10}~M_\odot$ as \citet{2022MNRAS.512.5339R} before their disc pre-selection. To facilitate direct comparison with observational studies, we also implemented a lower stellar mass limit of $10^{9}~M_\odot$, producing an extended sample of 2735 galaxies (we note that the larger sample is only used to demonstrate trends).

\subsection{Galaxy Orientation \& Shape Determination}
\label{sec:frame}

We adopt a detection strategy that avoids pre-selecting galaxies based on disk morphology. This agnostic approach allows us to probe the presence of triaxiality across the full morphological spectrum, extending previous TNG50 studies \citep[e.g.][]{2022MNRAS.512.5339R} that primarily focused on clear disk galaxies. 

To define a reproducible intrinsic frame for each galaxy, we determine the principal axes of the stellar mass distribution using an iterative reduced inertia tensor method, following the procedure outlined in  \citet{2018MNRAS.473.1489L}. The reduced inertia tensor $I_{ij}$ is computed to mitigate the bias from substructures in the outer halo while retaining sensitivity to the overall shape:
\begin{equation}
    I_{ij} = \sum_{k} \frac{x_{k, i} x_{k, j}}{R_{k, \text{ell}}^2},
\end{equation}
where $x_{k, i}$ is the $i$-th coordinate of the $k$-th stellar particle relative to the galaxy centre. The weighting factor is the squared ellipsoidal radius, defined as:
\begin{equation}
    R_{k, \text{ell}}^2 = x_k^2 + \frac{y_k^2}{q^2} + \frac{z_k^2}{s^2},
\end{equation}
where $x, y, z$ are aligned with the major, intermediate, and minor axes, respectively. The axis ratios $q = b/a$ and $s = c/a$ (with $a \ge b \ge c$) are initially set to unity (spherical symmetry). We calculate the tensor using stellar particles located within an ellipsoidal radius of $R_{k, \text{ell}} \le 5\,r_{\rm eff}$, where effective radius $r_{\rm eff}$ means half-mass radius. 

The diagonalization of $I_{ij}$ yields the eigenvectors (principal axes) and eigenvalues. We update the coordinate system and the axis ratios $q$ and $s$ iteratively until the variations between consecutive steps drop below 1 per cent. Finally, the galaxy is rotated into this converged principal-axis frame. We align the $z$-axis with the short axis ($c$) to define the ``face-on'' projection. We verified the robustness of this frame transformation by visually inspecting the projected stellar density maps for our sample in three orthogonal views after alignment, confirming that the resulting coordinate frame correctly aligns with the galaxy's visual major axes. Additionally, we ensure a consistent kinematic orientation by flipping the axes if necessary, such that the net stellar rotation within the aperture is counter-clockwise.

To determine the galaxy's primary rotation axis, we utilize the particle positions $\mathbf{x}$ and velocities $\mathbf{v}$, which have been pre-aligned with the principal axes of the inertia tensor. We first calculate the total angular momentum of the galaxy via $\mathbf{J} = \sum_i m_i (\mathbf{x}_i \times \mathbf{v}_i)$. The component of $\mathbf{J}$ with the largest magnitude indicates the geometric axis about which the galaxy primarily rotates; specifically, a maximum projection along the $x$, $y$, or $z$ axis corresponds to rotation about the major ($a$), intermediate ($b$), or minor ($c$) axis, respectively. To quantify the alignment between the kinematic and morphological axes, we compute the kinematic misalignment angle $\Psi$ relative to a specific geometric principal axis unit vector $\mathbf{\hat{e}}_k$ (e.g. $\mathbf{\hat{e}}_k = [0, 0, 1]^\mathrm{T}$ for the minor axis). In the principal axis frame, this is defined by $\cos \Psi = |\mathbf{J} \cdot \mathbf{\hat{e}}_k| / (\|\mathbf{J}\| \|\mathbf{\hat{e}}_k\|) = |J_k| / \|\mathbf{J}\|$. We note with such method, we find no clue of prolate rotators (rotating around their long axis, \citealp{2017A&A...606A..62T}).

\subsection{Bar-like structure Identification}

In this work, we use the term ``bar-like structures" to encompass both classical bars in discs and the elongated, rotating structures in quenched systems. We identify bar-like structures and characterize their properties by performing a Fourier decomposition on the face-on stellar mass surface density distribution. Following the standard methodology \citep{1986MNRAS.221..213A, 2002MNRAS.330...35A, 2013MNRAS.429.1949A}, we quantify the bar strength using the normalised amplitude of the bisymmetric ($m=2$) Fourier mode, defined as:
\begin{equation}
    A_2(R) = \frac{\left| \sum_j m_j \exp(2i\theta_j) \right|}{\sum_j m_j},
\end{equation}
where $m_j$ and $\theta_j$ represent the mass and azimuthal angle of the $j$-th star particle, respectively, and the summation extends over all particles within a given radial bin at radius $R$.

To reliably detect bars and determine their physical extent, we adopt a procedure basically based on the criteria of \citet{2022MNRAS.512.5339R}. Our identification pipeline proceeds (while we use stricter selection criteria) as follows:
(i) \textit{Strength threshold:} We first search for a significant peak in the radial $A_2(R)$ profile, requiring a maximum amplitude of $A_{2,\rm peak} > 0.2$.
(ii) \textit{Bar length:} The radial extent of the bar is determined by tracing the $A_2$ profile outwards from the peak until the amplitude drops below either $0.5\,A_{2,\rm peak}$ or a floor value of $0.15$.
(iii) \textit{Phase coherence:} To distinguish bars from spiral arms, we require the phase of the $m=2$ mode, $\phi_2(R)$, to remains nearly constant across the bar region. Specifically, the phase variation must satisfy $|\Delta \phi_2| \le 10^\circ$ within the identified radius.
(iv) \textit{Minimum length:} We classify a galaxy as barred only if the determined bar length exceeds $1.2~\mathrm{kpc}$.

We have verified that the sample of barred galaxies obtained via this pipeline shows complete agreement (100\%) with the catalog of \citet{2022MNRAS.512.5339R} when applied to their parent disk galaxy ($D/T>0.5$) sample. Finally, to ensure the robustness of our face-on projection, galactic center centering, and bar detection, we performed a manual visual inspection of the three orthogonal projections for every galaxy in our sample.

The pattern speed of the identified bars is calculated by applying the method of \citet{2023MNRAS.518.2712D} to individual simulation snapshots, which ensures high precision. We also adopt the concept of ``local pattern speed'' introduced in our companion paper \citep{2026arXiv260305287D} to examine the radial profile of the pattern speed. This allows us to verify whether the pattern speed exhibits a constant plateau within the bar region and clearly decouples from the stellar streaming motion, which are characteristic signatures of a classical density wave.

\subsection{Disk Property Determination}

To quantify the structural concentration of the galaxies, we compute the concentration index defined as the ratio $R_{90}/R_{50}$ \citep{2010ApJ...714L.260N, 2025MNRAS.542..151M}. For a single galaxy, its stellar particles are projected onto the $xy$-plane to derive the two-dimensional radial distribution. To construct the light profile, we calculate the cumulative flux distribution weighted by the $r$-band luminosity. $R_{90}$ and $R_{50}$ are then determined as the projected radii enclosing 90 per cent and 50 per cent of the total integrated flux respectively. 

To characterize the kinematic morphology, we adopt the kinematic disc-to-total ratio ($D/T$). Following the methodology in \citet{2022MNRAS.512.5339R, 2020ApJ...895...92Z}, galaxies are re-oriented to align the total stellar angular momentum with the $z$-axis. For all stellar particles within $3 \, r_{\rm eff}$ (half-mass radius), we compute the circularity parameter $\epsilon = J_z / J(E)$, where $J_z$ is the specific angular momentum along the symmetry axis and $J(E)$ is the maximum specific angular momentum at the particle's binding energy $E$. We define ``disc stars'' as those with $\epsilon > 0.7$. The $D/T$ ratio is then calculated as the mass fraction of these disc stars relative to the total stellar mass in the considered volume. While \citet{2022MNRAS.512.5339R} restricted their analysis to a disc-dominated parent sample (requiring $D/T \geq 0.5$), we do not impose such pre-selection. Instead, we identify bars, or say bar-like structures from the total galaxy population and utilize $D/T$ as a metric to quantify the kinematic state of the host systems. 

To provide a metric comparable to observations, we derive the S\'ersic index ($n$) following the ``mock-ETG" pipeline of \citet{2025MNRAS.539.2855D}. 
After projecting stellar particles along a principal axis, we measure the axis ratio ($q\equiv c/a$) and position angle from the light-weighted moment of inertia tensor within a 5\,$r_{\rm eff}$ aperture. 
We adopt a circularized radius $r_{\rm circ} \equiv \sqrt{q x^2 + y^2/q}$ (aligned to the major axis) to compute the cumulative light profile and obtain the non-parametric half-light radius $R_{e,\mathrm{dir}}$. 
The surface brightness profile $I_{\rm obs}(r)$ is then extracted in 40 logarithmic annuli over the range $[\,\max(1\,\mathrm{kpc},\, f R_{e,\mathrm{dir}}),\, 3R_{e,\mathrm{dir}}\,]$, where the inner boundary factor adapts to resolution ($f=0.2$ for $R_{e,\mathrm{dir}} \geq 10$\,kpc, and $0.35$ otherwise). 
We determine $n$ by fitting a standard S\'ersic law $I(r)=I_e\exp\{-b_n[(r/R_e)^{1/n}-1]\}$ via robust non-linear least squares (using a soft-$L_1$ loss to handle outliers) in $\log_{10}I$ space, adopting the \citet{2003ApJ...582..689M} approximation for $b_n(n)$.

\section{Results} \label{sec:results}

We report the statistics of bar-like structures in TNG50 galaxies at $z=0$. Unlike observational censuses, which typically pre-select galaxies based on disk morphology, we analyze the full mass-limited sample. This morphology-agnostic approach enables us to determine whether bar-like instabilities—or triaxial structures identified as bars via Fourier decomposition—are truly exclusive to disk galaxies.

\begin{figure*}
	\centering
    % Row 1: b/a
	\includegraphics[width=0.9\textwidth]{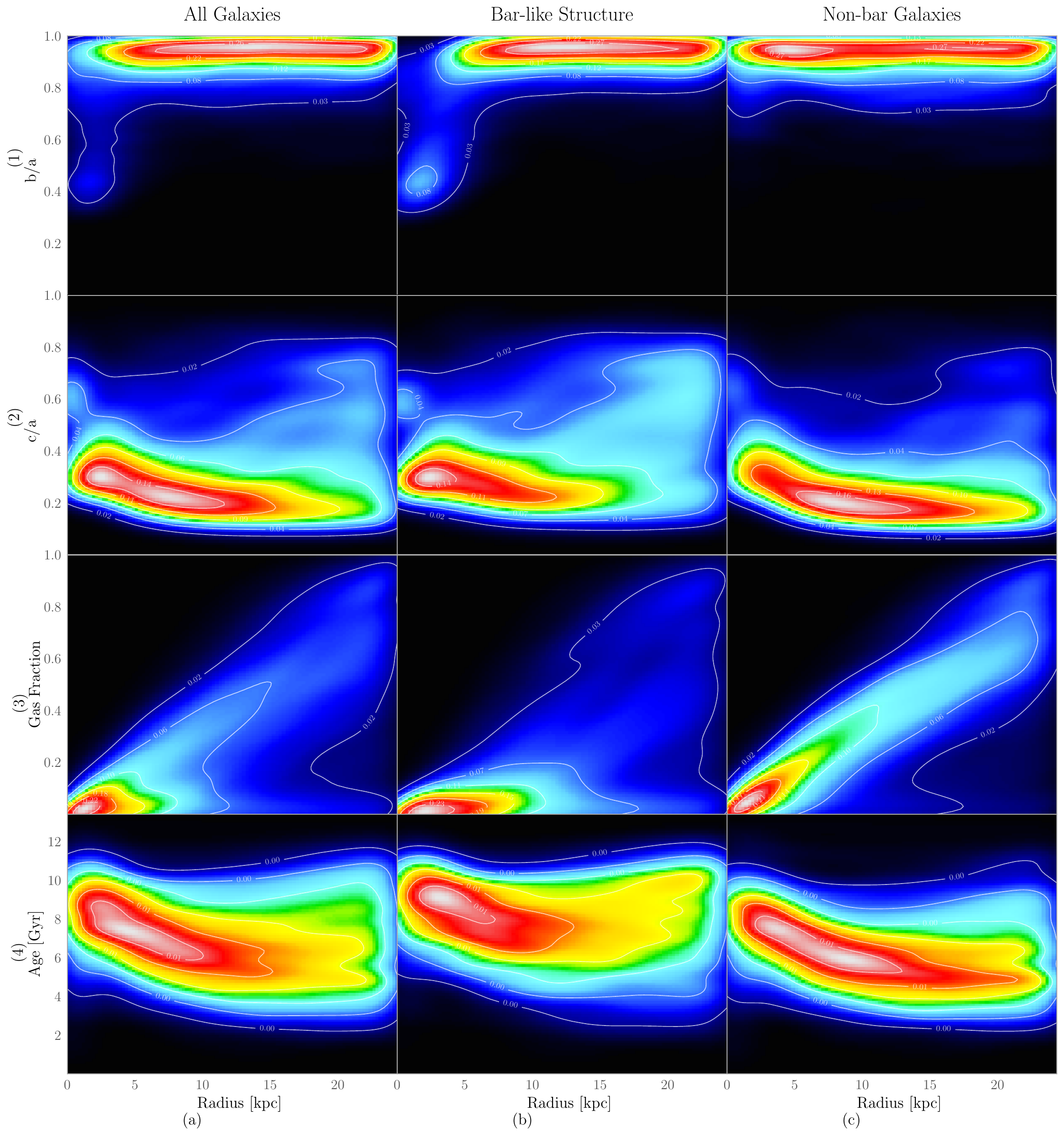}
    
	\caption{
    \textbf{Stacked radial profiles of galaxy properties in TNG50.} 
    Each panel displays the KDE-smoothed density of stacked radial profiles for the axis ratios $b/a$ (Row 1) and $c/a$ (Row 2), gas fraction (Row 3), and stellar age (Row 4). 
    The columns demarcate the sample: \textbf{Left (a):} All galaxies; \textbf{Center (b):} Galaxies identified with bar-like structures; \textbf{Right (c):} Non-barred galaxies.
    \textbf{Row 1 ($b/a$):} A distinct bimodality emerges in the inner $<4$ kpc (Panel 1a), identifying the bar population.
    \textbf{Row 2 ($c/a$):} The central overdensity at $c/a > 0.5$ ($<2$ kpc) represents classical bulges. Comparison of 2b and 2c reveals that bar-like structures are more likely to be bulge-dominated than non-barred galaxies, consistent with Galaxy Zoo 2 \citep{2012MNRAS.423.1485S}. What is more, Panel 2b exhibits a significantly different overdensity compared to Panel 2c in the outer $>10$ kpc.
    \textbf{Rows 3 \& 4;} Bar-like structures are systematically associated with gas-poor (Panel 3b) and older (Panel 4b) stellar populations compared to the non-barred control sample.
    }
	\label{fig:stacked_profiles}
\end{figure*}

\subsection{The Morphological Distinctness of Bar-like Structures} \label{sec:morphology_stack}

Before quantifying the statistics of bar-like structures, we first characterize their morphology and connection to the host galaxy's structural evolution inside the $M_\star>10^{10}~M_\odot$ sample. We achieve this by ``stacking'' the radial property profiles of our sample. For every galaxy, we compute the radial profiles of the axis ratios ($b/a$, $c/a$), gas fraction, and stellar age. These profiles are then superimposed to construct a two-dimensional histogram, smoothed via Kernel Density Estimation (KDE) to visualize the global structural trends of the population.

Figure~\ref{fig:stacked_profiles} presents these stacked landscapes. The rows correspond to the evolution of $b/a$, $c/a$, gas fraction, and stellar age with radius, while the columns separate the sample into the total population, bar-like-structured systems, and non-barred systems.

\subsubsection{Face-on Profiles $b/a$}

The minor-to-major axis ratio ($b/a$) profiles (Row 1) reveal a feature: the TNG50 galaxy population exhibits a natural, bimodal segregation in the inner galactic regions ($R < 4$ kpc). The total population (Panel 1a) clearly bifurcates into a near-circular regime ($b/a \sim 1$) and a horizontally elongated density enhancement ($b/a \sim 0.4$). By separating the sample, we confirm that this elongated overdensity corresponds precisely to our identified bar-like structures (Panel 1b), while non-barred galaxies remain circular (Panel 1c). 

It is worth noting that the location of this overdensity agrees with the $b/a < 0.6$ cut employed by \citet{2020A&A...641A..60P} to remove inconsistent objects, and the selection criteria used by \citet{2021A&A...647A.143L} to flag bar-like galaxies. Furthermore, in the galaxy outskirt regime ($R > 5$ kpc), the distribution of $b/a$ is unimodal and continuous close to $1$, suggesting the oblate nature of most galaxies. 

\subsubsection{Vertical Structure $c/a$}

The vertical axis ratio ($c/a$) profiles (Row 2) provide insight into the thickness and bulge properties. In the total population (Panel 2a), an overdensity appears at $R < 2$ kpc with $c/a > 0.5$, indicative of classical bulges. Breaking the whole sample down by the presence of bar-like structures and bulges reveals a tripartite structural classification. Non-barred galaxies (Panel 2c) are almost exclusively thin, bulgeless systems. Conversely, the bar-like-structure population (Panel 2b) hosts both systems with prominent central spheroids (bar+bulge) and those without (bar+pseudo-bulge). Compared to non-barred galaxies, the bar-like-structure population exhibit a higher prevalence to be bulge-dominated, consistent with Galaxy Zoo 2 \citep{2012MNRAS.423.1485S}.

The ($c/a$) analysis for bar-like-structure galaxies (Panel 2b) exhibits a significantly different overdensity compared to Panel 2c in the outer $>10$ kpc. Notably, a significant fraction of these objects possess $c/a > 0.6$, indicating bar-like structures that extend to large radii with negligible disc components. 

\subsubsection{Gas and Age: The Quenching Connection}

Finally, the gas fraction and stellar age profiles (Rows 3 and 4) confirm that these structural differences are intimately linked to the evolutionary state of the galaxy. The bar-like population is systematically gas-poor (Panel 3b vs 3c) and composed of older stellar populations (Panel 4b vs 4c) at all radii. This reinforces the picture that the ``excess'' bar-like structures in TNG50 are not random features but are deeply embedded in the physics of quenching system. TNG50 suggests that as galaxies consume their gas and age, they tend to settle into the specific region of morphological phase-space defined by central elongation ($b/a \sim 0.4$) and vertical thickening.

\subsection{Tension and Consistency with Observations} \label{sec:tension_consistency}

We now turn to the critical question: how do the statistics of these bar-like structures in TNG50 compare with the real Universe? A direct comparison is non-trivial. Observational bar fractions are widely defined as $f_{\rm bar} = N_{\rm barred} / N_{\rm disc}$, derived from samples rigidly pre-selected to exclude elliptical galaxies and merger remnants based on colour, concentration, or visual morphology. In contrast, our analysis adopts a morphology-agnostic definition, $f_{\rm bar} = N_{\rm barred} / N_{\rm total}$, to probe the intrinsic prevalence of these structures without imposing a priori assumptions about their host galaxies.

Comparing these inherently different parent samples is nonetheless scientifically instructive. By contrasting the intrinsic population predicted by TNG50 against the pre-filtered populations in observational censuses, we can isolate the impact of selection effects. Specifically, where the two statistics diverge, it highlights regimes where TNG50 predicts the existence of bar-like structures in galaxies that observational criteria would typically classify as ``unbarred ETGs'' or ``featureless spheroids'' and thus discard from the analysis.

\begin{figure*}
	\centering
	\begin{tabular}{cc}
    % Row 1: Stellar Mass / Kinematics
		\includegraphics[width=0.45\textwidth]{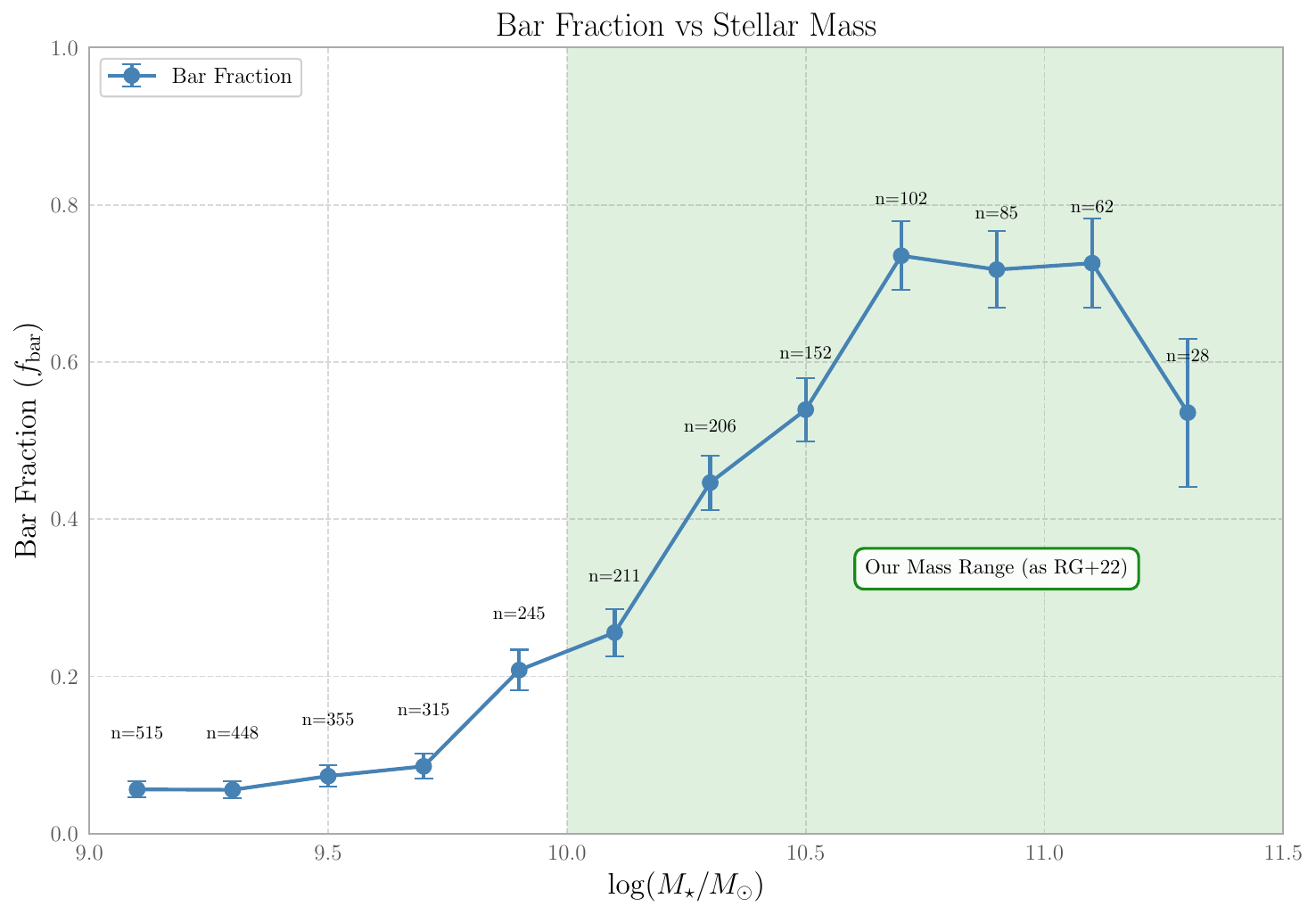} &
		\includegraphics[width=0.52\textwidth]{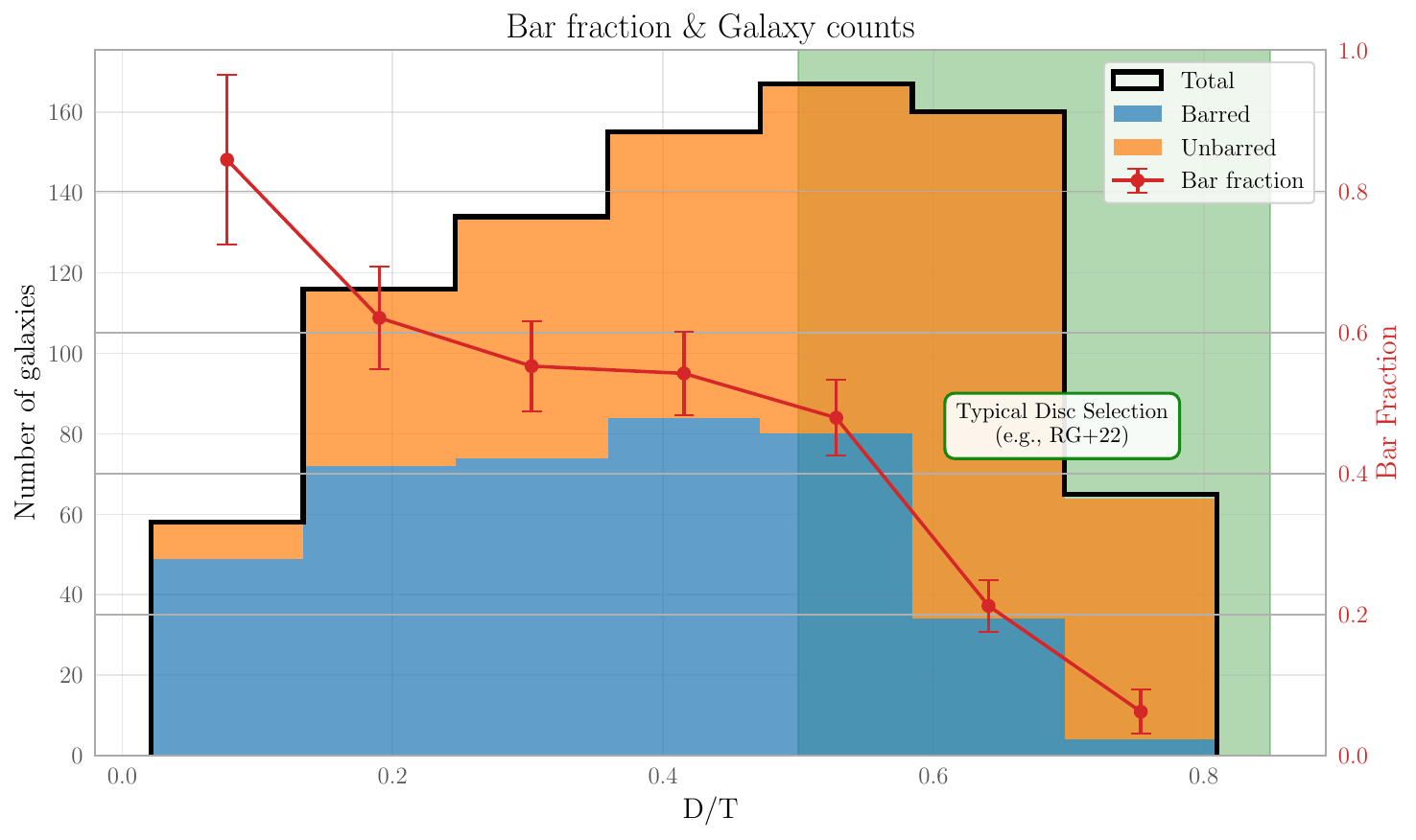} \\
    (a) $f_{\rm bar}$ vs. Stellar Mass & (d) $f_{\rm bar}$ vs. Kinematics ($D/T$) \\[1em]
    % Row 2: Colour / Sersic Index
		\includegraphics[width=0.45\textwidth]{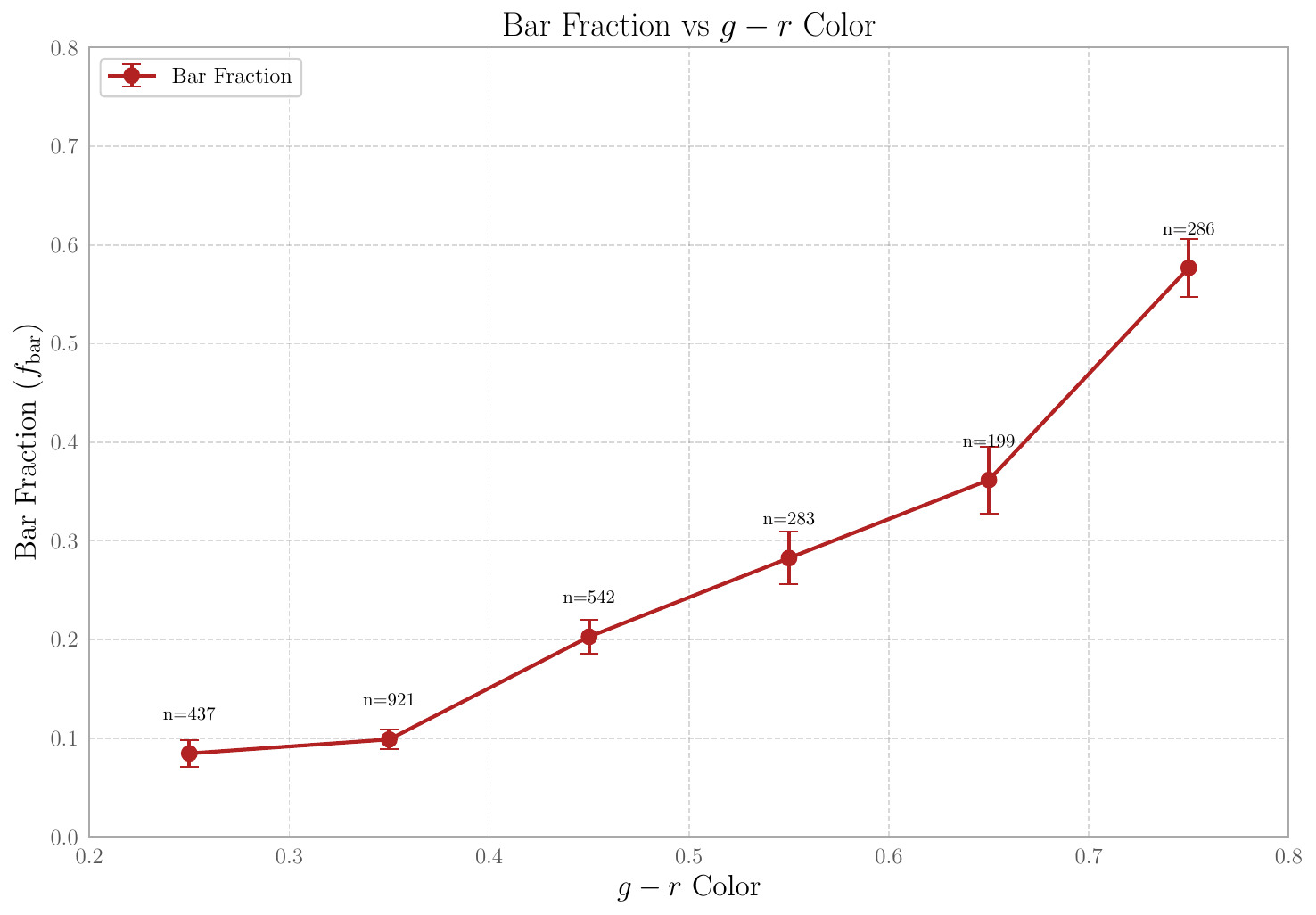} &
		\includegraphics[width=0.52\textwidth]{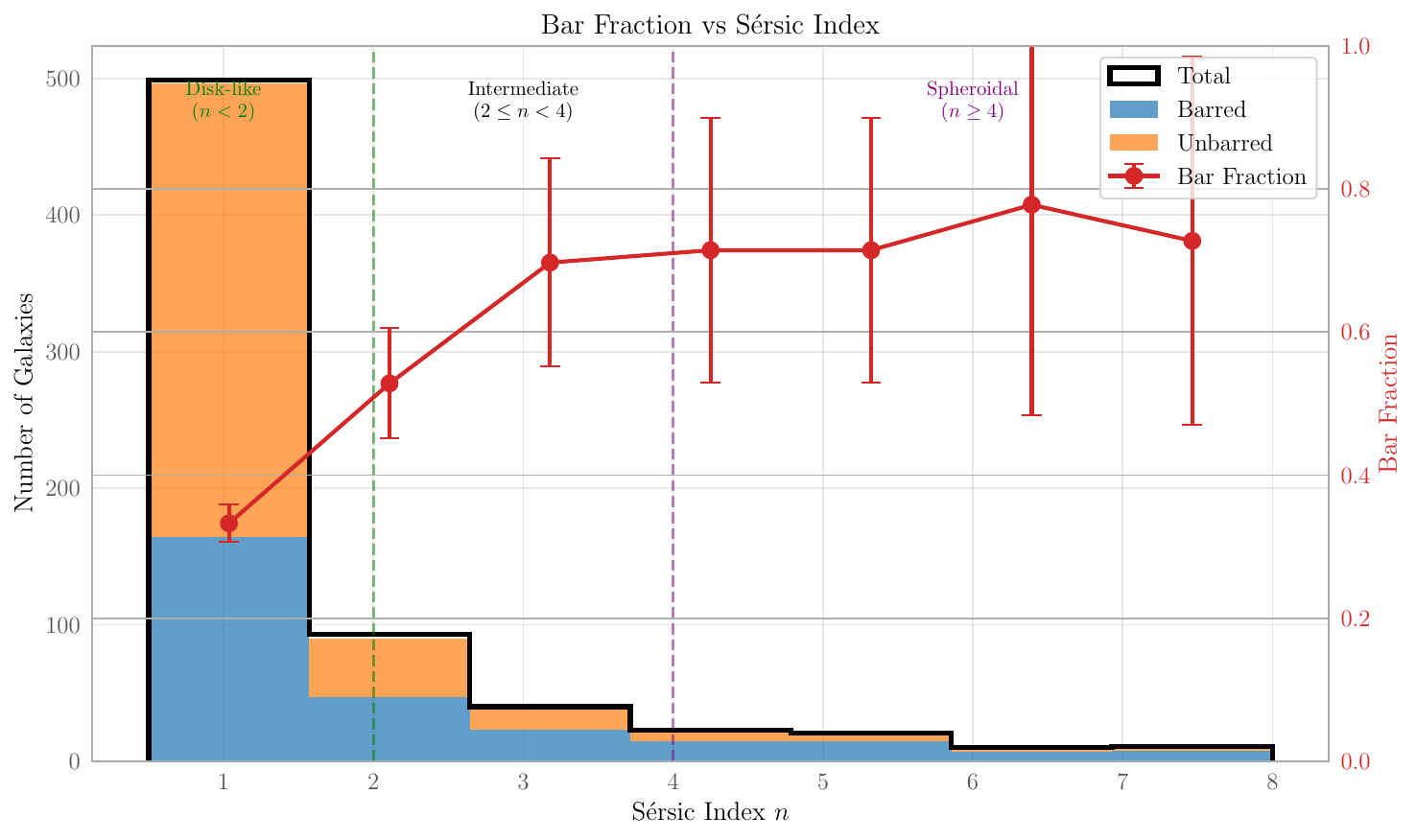} \\
    (b) $f_{\rm bar}$ vs. Colour & (e) $f_{\rm bar}$ vs. S\'ersic Index \\[1em]
    % Row 3: Concentration / Bar Properties
		\includegraphics[width=0.44\textwidth]{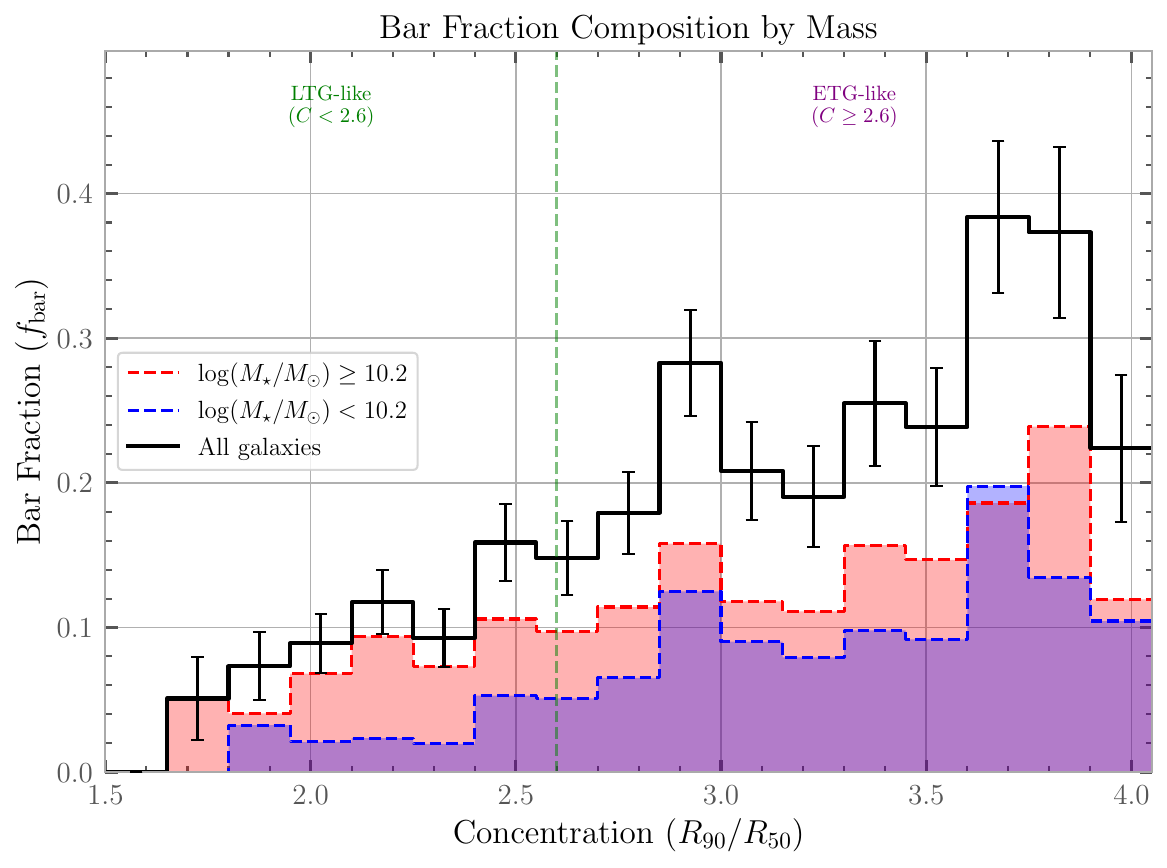} &
		\includegraphics[width=0.52\textwidth]{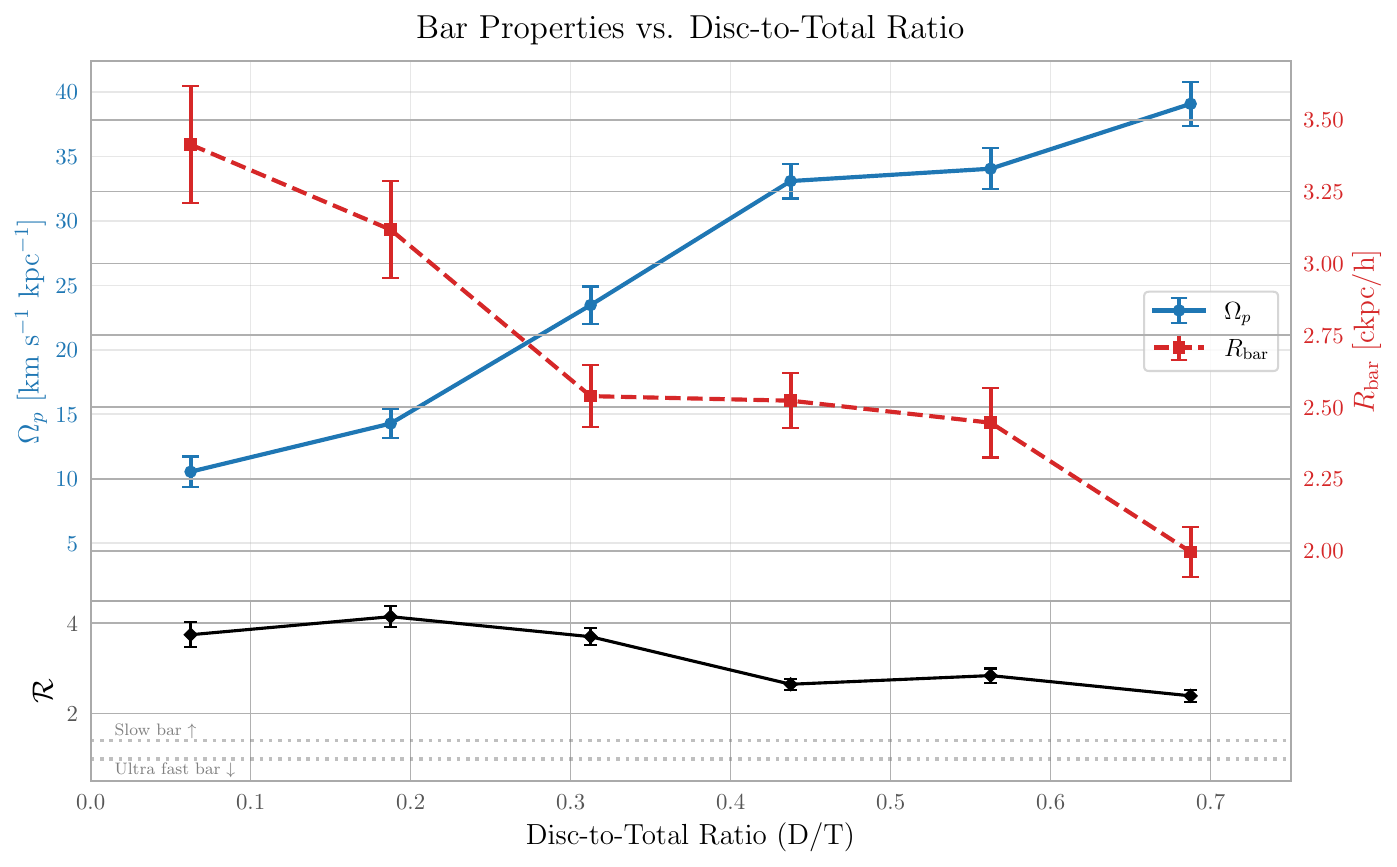} \\
    (c) $f_{\rm bar}$ vs. Concentration & (f) bar properties vs. $D/T$ \\
	\end{tabular}
	\caption{The demographics of bar-like structures in TNG50 at $z=0$. \textbf{Left Column:} For comparing with observations. (a) Bar fraction as a function of stellar mass. The green shaded region marks the mass range ($>10^{10} \rm M_{\odot}$) used in previous TNG studies \citep{2022MNRAS.512.5339R}. (b) Bar fraction vs. $g-r$ colour, showing a monotonic increase towards redder systems. (c) Bar fraction vs. concentration index $R_{90}/R_{50}$ for different mass cuts, highlighting the high prevalence of bar-like structures in highly concentrated (ETG-like) systems. \textbf{Right Column:} Internal structural dependencies for comparing with previous TNG works (d) Bar fraction vs. kinematic disc-to-total ratio ($D/T$). The green zone indicates typical disc selection cuts. (e) Bar fraction vs. S\'ersic index $n$. $n$ can be used for ETG  classification \citep{2025MNRAS.539.2855D}. (f) Mean pattern speed $\Omega_p$, bar length $R_{\rm bar}$ and ratio $\mathcal{R}\equiv R_{\rm CR}/R_{\rm bar}$as a function of $D/T$, showing that ``ETG bar-like structures" are physically distinct: longer and slower. Moreover, the ratio $\mathcal{R}$ is significantly larger than observed values, which was also reported by \citet{2022ApJ...940...61F}. Note that Panels~(c) and~(e) display the bar fraction and absolute counts, respectively. Although histograms conventionally denote counts, this format is used in Panel~(c) to ensure consistency with prior observational literature \citep{2010ApJ...714L.260N, 2025MNRAS.542..151M}.}
	\label{fig:obs_comparison_grid}
\end{figure*}

Figure~\ref{fig:obs_comparison_grid} presents the dependence of the bar fraction ($f_{\rm bar}$) on stellar mass, colour, and concentration (left column) for comparison with observations ($M_\star>10^{9}~M_\odot$ sample), alongside internal dependencies on kinematics and photometric profile shape (right column) for comparison with previous TNG works ($M_\star>10^{10}~M_\odot$ sample).

\subsubsection{Mass, Colour, and Concentration Trends}
\label{sec:compare_with_observation}

In Panel (a), we examine $f_{\rm bar}$ as a function of stellar mass, extending the lower limit to $M_* \ge 10^{9} \rm M_{\odot}$. The distribution is unimodal, peaking at $\sim 75\%$ for $M_* \approx 10^{10.8} \rm M_{\odot}$ before declining at the highest masses. At the low-mass end ($<10^{9.5} \rm M_{\odot}$), the fraction drops to $\sim 5\%$, indicating that the global bar fraction is highly sensitive to the sample mass floor. This trend and peak aligns with results from Galaxy Zoo Hubble \citep[GZH;][Fig.5]{2014MNRAS.438.2882M} and $S^4G$ \citep[][Fig.19]{2016A&A...587A.160D}. 

Panel (b) reveals a strong positive correlation between bar fraction and galaxy colour (derived from the TNG official \texttt{SubhaloStellarPhotometrics} dataset): redder galaxies are significantly more likely to look barred. This trend shows agreement with the citizen-science classifications from Galaxy Zoo 2 \citep[GZ2;][Fig.3]{2011MNRAS.411.2026M}. 

We note studies using SDSS \citep[][Fig.4]{2010ApJ...714L.260N}, DESI/ALFALFA \citep[][Fig.~6]{2025MNRAS.542..151M}, and SDSS/MaNGA \citep[][Fig.8]{2022MNRAS.512.2222V, 2026MNRAS.546ag015A} have reported bimodal distributions in the $f_{\rm bar}-M_\star$ and $f_{\rm bar}-$color planes. Crucially, while the early-type populations (S0--Sb; concentration $>2.4$) aligns remarkably well with the trends we find in TNG50, the late-type populations (Sbc--Sm; concentration $<2$) display a completely opposite correlation.

We plot $f_{\rm bar}$ against the concentration index $C = R_{90}/R_{50}$ in Panel (c), where $f_{\rm bar}$ ascends with $C$ in all sub-samples monotonically (observationally, galaxies with $C > 2.6$ are typically dominated by ETGs, \citealp{2001AJ....122.1861S}). For comparison, the more massive sub-sample in SDSS \citep[][Fig.5]{2010ApJ...714L.260N} and DESI/ALFALFA \citep[][Fig.6]{2025MNRAS.542..151M} both agree with such trend, while the less massive sub-sample shows the opposite behavior. This causes the full sample in both papers to exhibit bimodal/monotonically decreasing trends respectively.

What is more, $S^4 G$ (\citealp[][Fig.7]{2015ApJS..217...32B}; \citealp[][Fig.10]{2016A&A...587A.160D}) and SDSS/MaNGA \citep[][Fig.8]{2022MNRAS.512.2222V} have shown bimodal distributions in Hubble type, where bar fractions can possibly rise in early-type bins.

\subsubsection{Structural and Kinematic Properties}

The bottom row of Figure~\ref{fig:obs_comparison_grid} helps to  compare with previous works on IllustrisTNG.

Panel (d) shows the relationship with kinematic morphology ($D/T$). Previous TNG studies have focused on the regime $D/T \gtrsim 0.5$ (green shaded area), finding moderate bar fractions consistent with observations \citep{2022MNRAS.512.5339R}. However, outside this window, in the dispersion-dominated regime ($D/T < 0.2$), the bar fraction surges to $\sim 80\%$. A similar trend is visible in Panel (e) using the photometric S\'ersic index: while disc-like systems ($n < 2$) have modest bar fractions ($\sim 35\%$), spheroidal systems ($n \ge 4$) are overwhelmingly barred ($\sim 75-80\%$). We note S\'ersic index is one of the factors for ETG selection in \citet{2025MNRAS.539.2855D}.

This confirms that strong triaxiality is a near-universal feature of simulated massive ETGs in TNG50. While often classified as ``prolate'' in shape analyses, our kinematic analysis confirms these are rotating structures around short axis not long axis. Indeed, recent observational work with MaNGA \citep{2026MNRAS.546ag015A} has detected a similar inverse correlation between bar fraction and stellar angular momentum parameter $\lambda_R$ , hinting that nature may indeed hide bars in low-spin systems.

Finally, Panel (f) characterizes the bars (bar-like structures) against kinematic $D/T$. We observe a clear dichotomy: as galaxies become more dispersion-dominated (lower $D/T$), their bars become physically \textit{longer} and rotate significantly \textit{slower}. This implies that the ``excess'' bar-like structures in TNG50 ETGs are not merely numerical artifacts or transient features. They represent a population of highly evolved, slow-rotating bar-like structures embedded in hot stellar components. Moreover, TNG50 yields values of the parameter $\mathcal{R} \equiv R_{\mathrm{CR}}/R_{\mathrm{bar}}$ that significantly exceed observational benchmarks, placing the population deep within the slow bar regime ($\mathcal{R} > 1.4$).
As demonstrated by \citet{2022ApJ...940...61F}, the simulated pattern speeds (and thus $R_{\mathrm{CR}}$) are consistent with MaNGA observation, TNG50 bars are systematically shorter than observed \citep[see also][]{2020ApJ...904..170Z}; this underestimation of the denominator $R_{\mathrm{bar}}$ inflates $\mathcal{R}$.

\subsection{Anatomy of the Barred Population} \label{sec:case_studies}

The statistical results presented in Section~\ref{sec:results} suggest a continuum of bar-like structures extending from cold discs into the dispersion-dominated regime. To understand the physical validity of this continuity—and particularly to determine whether the structures identified in ETGs represent genuine bar instabilities versus numerical noise or transient triaxiality—we perform a detailed comparative anatomy of two representative systems.

\begin{figure*}
	\centering
    % Row 1: LTG bar ID 585282
    \begin{minipage}[c]{0.65\textwidth}
        \centering
        \includegraphics[width=\textwidth]{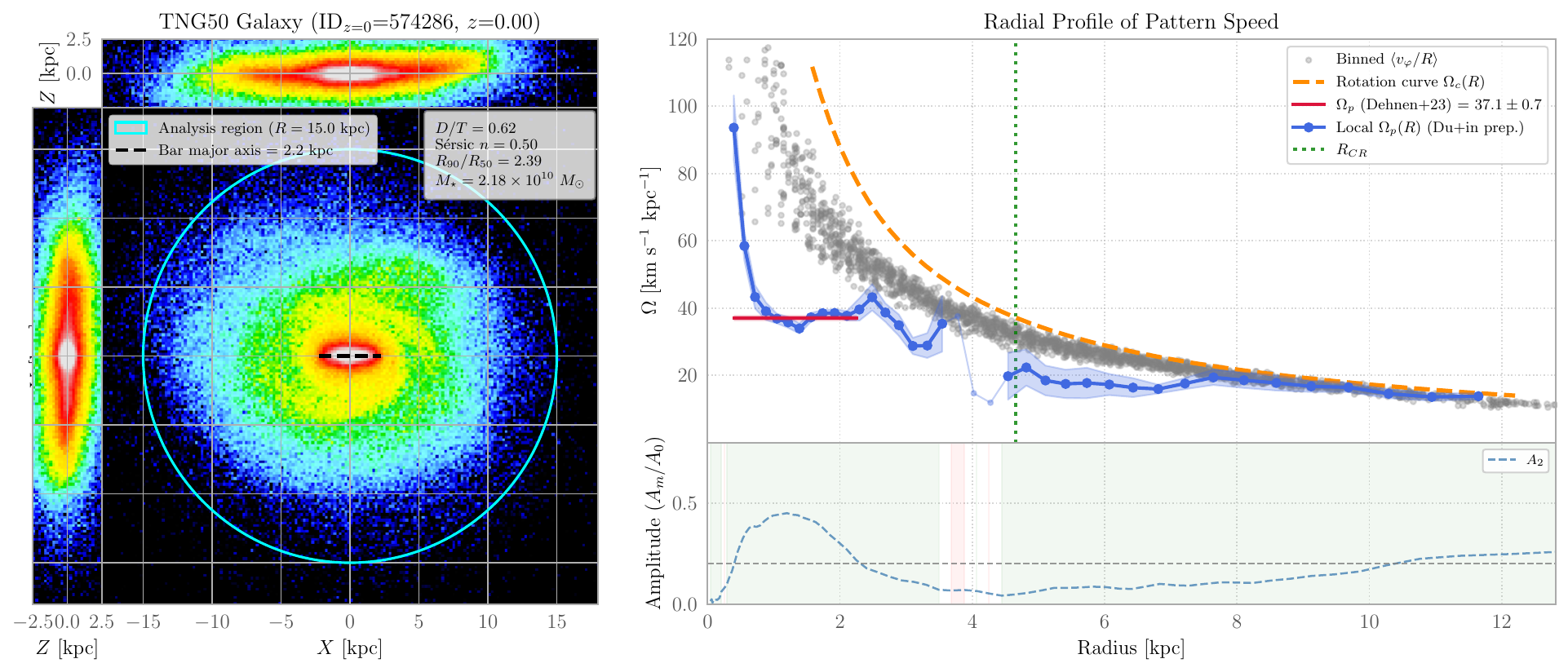}
    \end{minipage}%
    \begin{minipage}[c]{0.25\textwidth}
        \centering
        \includegraphics[width=\textwidth]{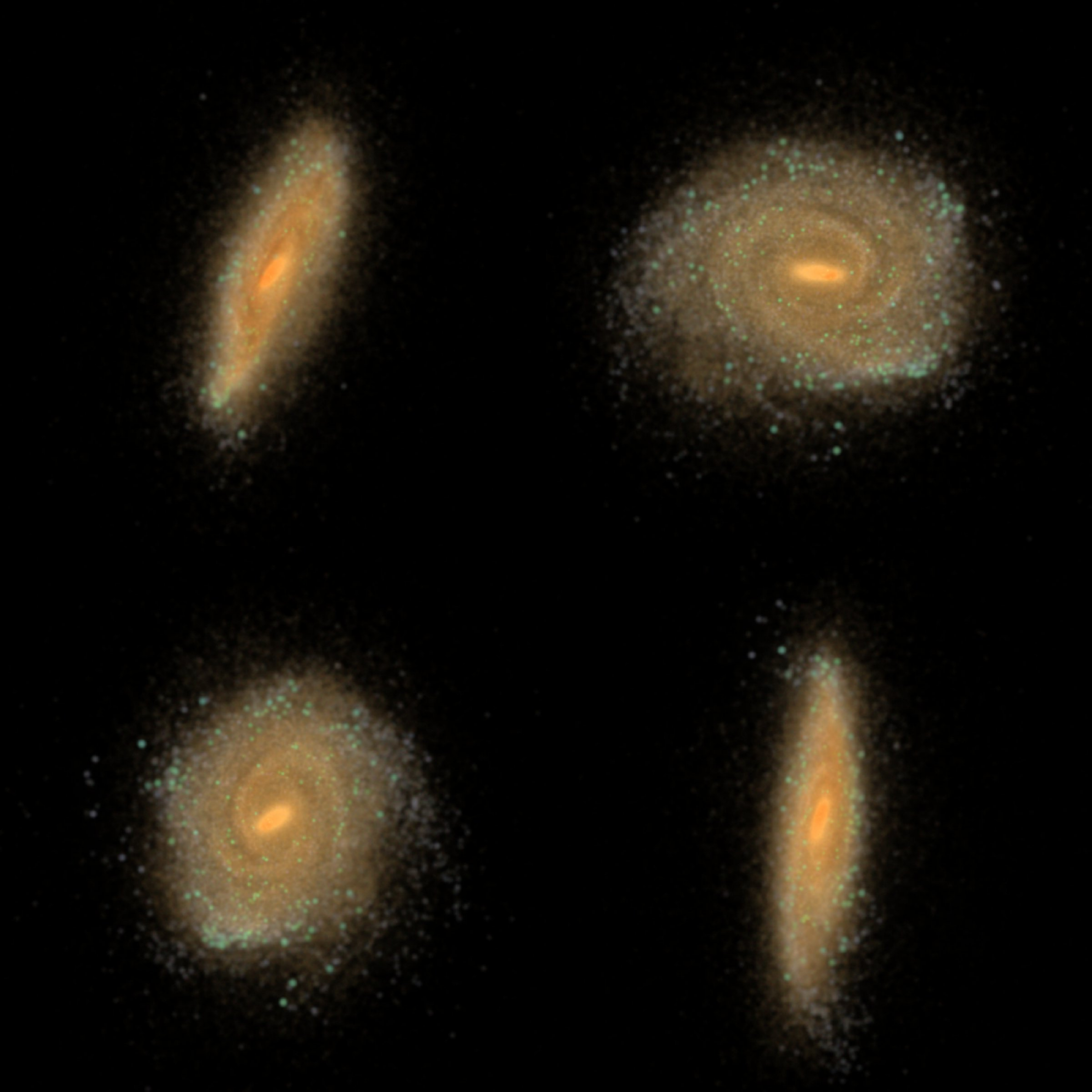}
    \end{minipage}
    \\[-0.6em]
    % Row 2: ETG bar ID 26
    \begin{minipage}[c]{0.65\textwidth}
        \centering
        \includegraphics[width=\textwidth]{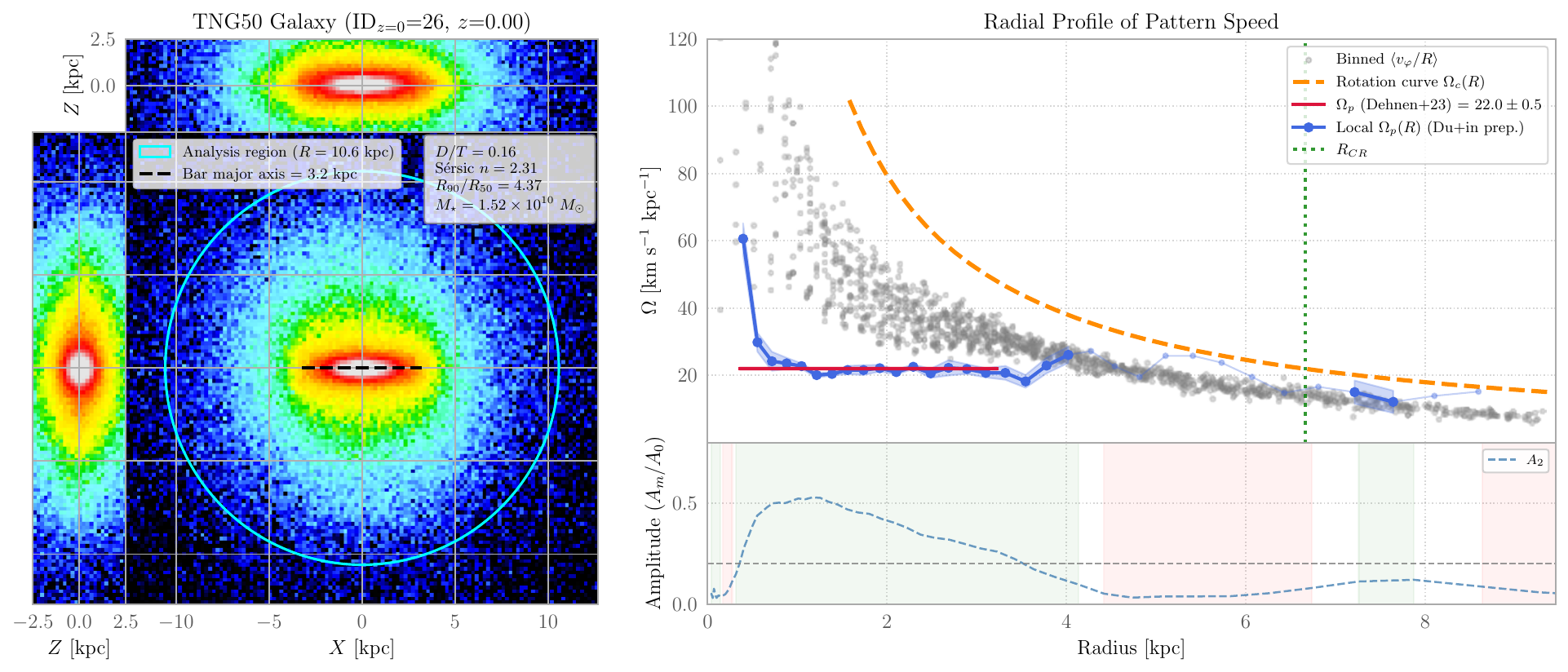}
    \end{minipage}%
    \begin{minipage}[c]{0.25\textwidth}
        \centering
        \includegraphics[width=\textwidth]{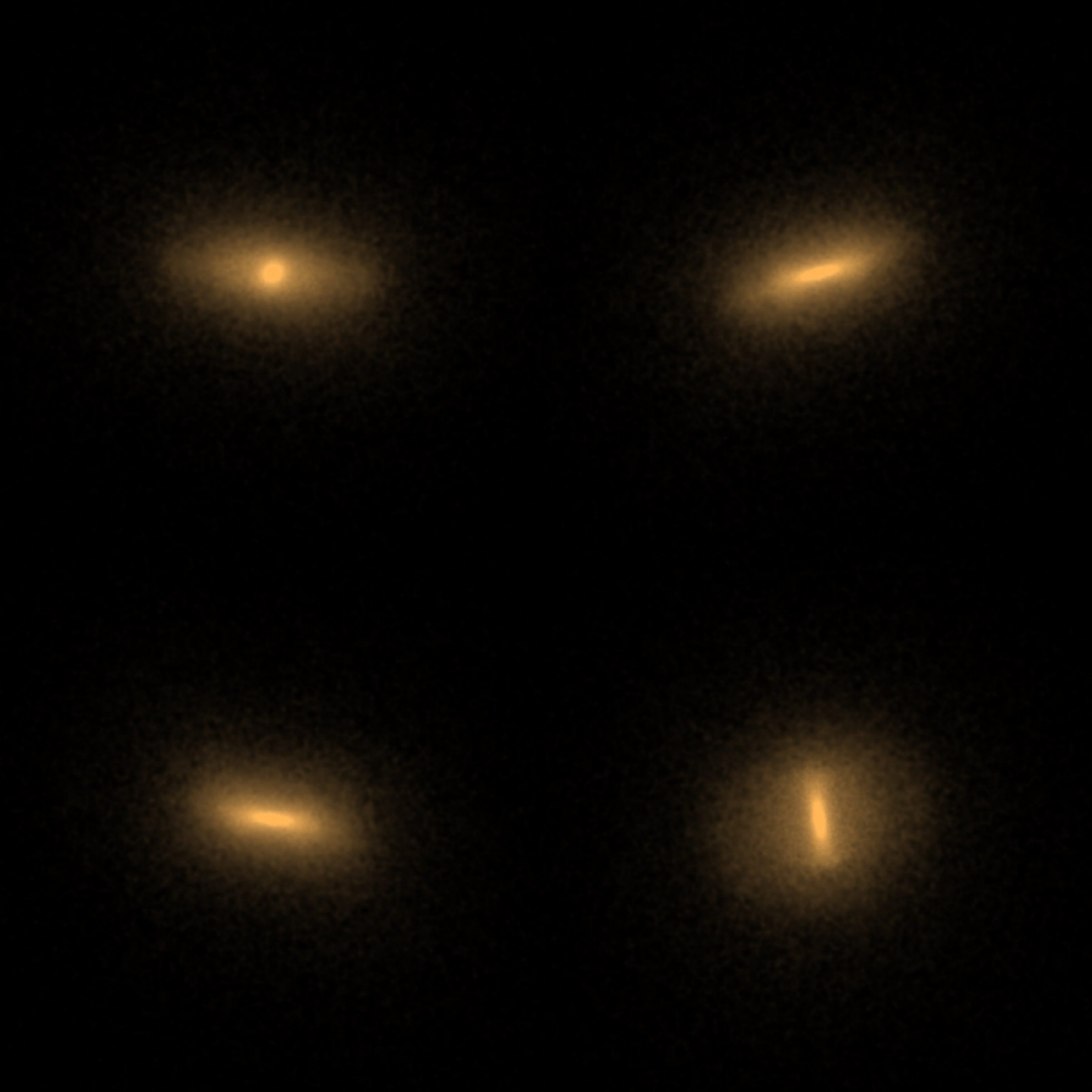}
    \end{minipage}
    \\[-0.6em]
    % Row 3: LTG unbar
    \begin{minipage}[c]{0.65\textwidth}
        \centering
        \includegraphics[width=\textwidth]{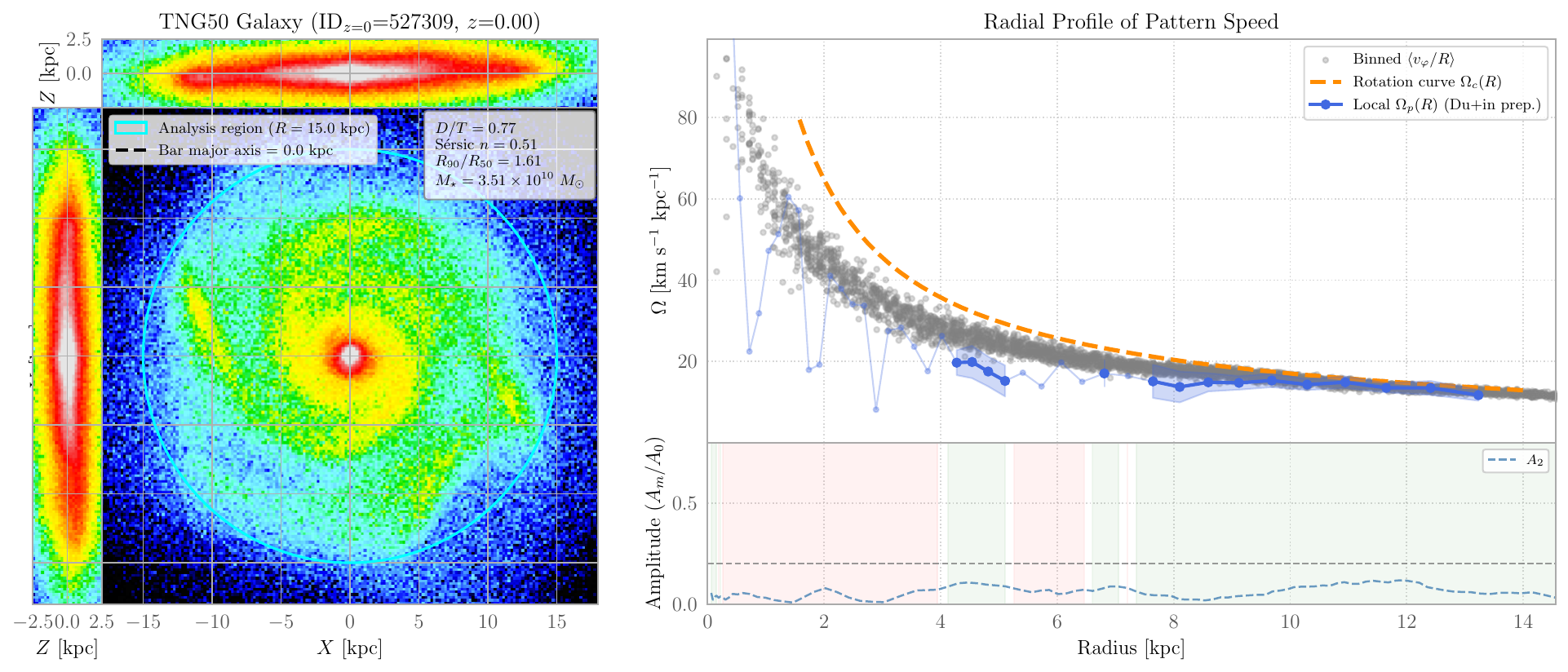}
    \end{minipage}%
    \begin{minipage}[c]{0.25\textwidth}
        \centering
        \includegraphics[width=\textwidth]{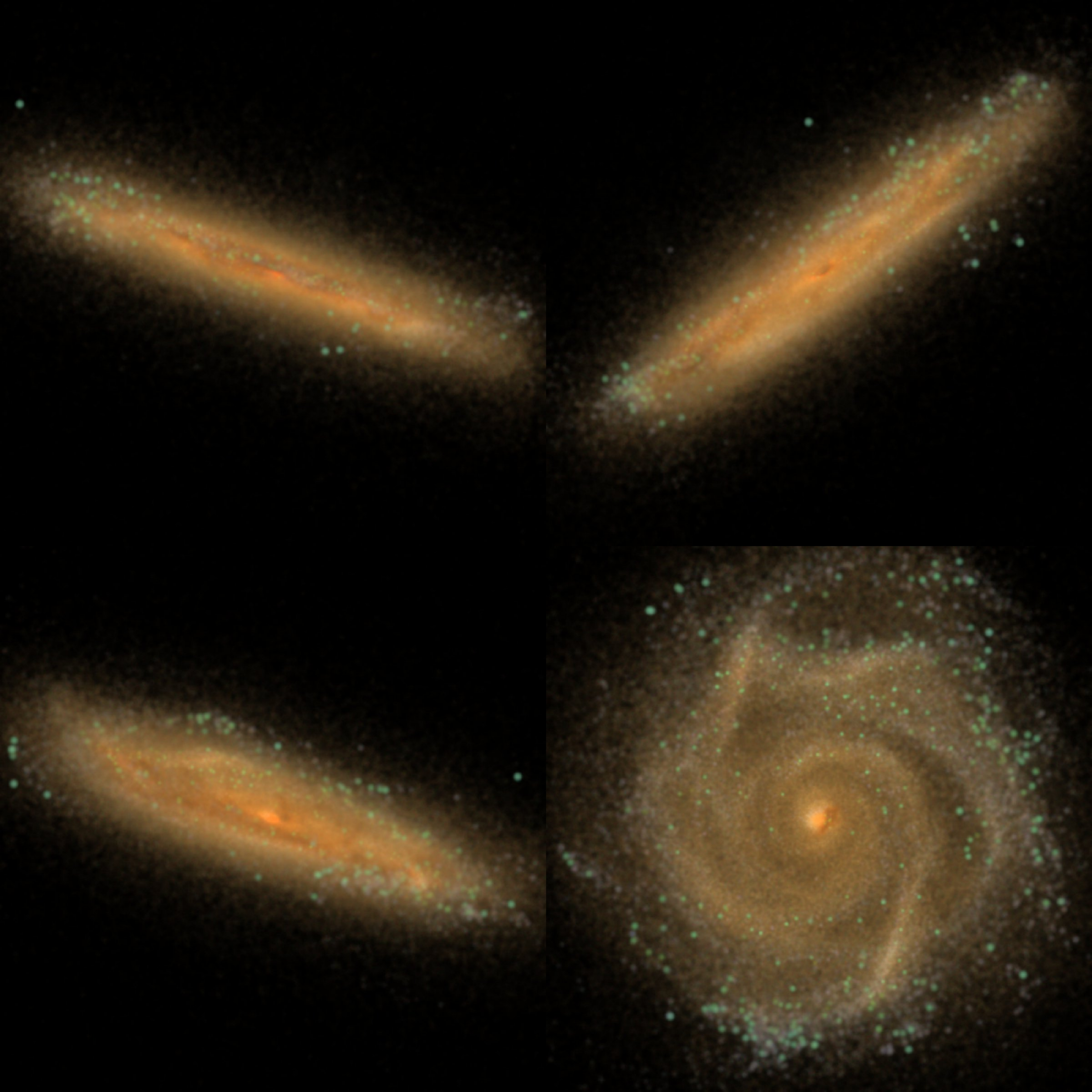}
    \end{minipage}
    \\[-0.6em]
    % Row 4: ETG unbar
    \begin{minipage}[c]{0.65\textwidth}
        \centering
        \includegraphics[width=\textwidth]{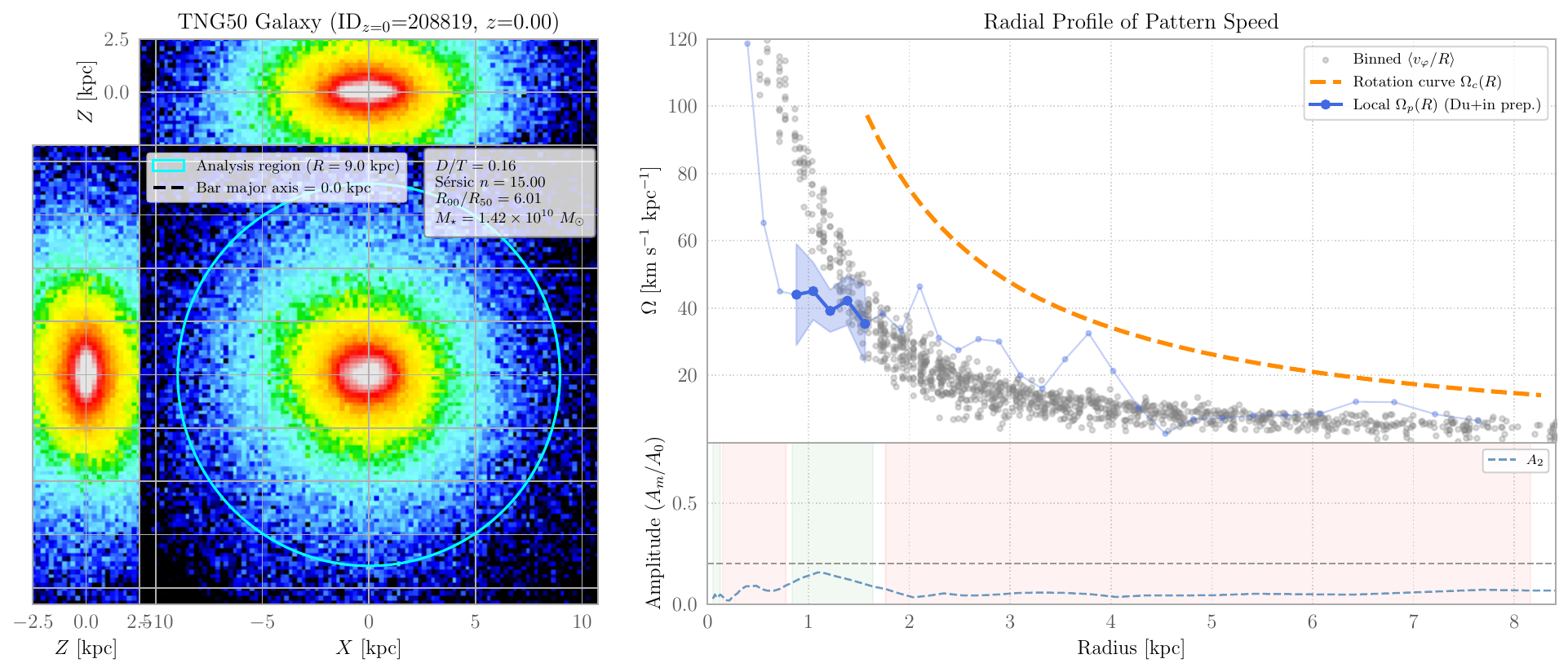}
    \end{minipage}%
    \begin{minipage}[c]{0.25\textwidth}
        \centering
        \includegraphics[width=\textwidth]{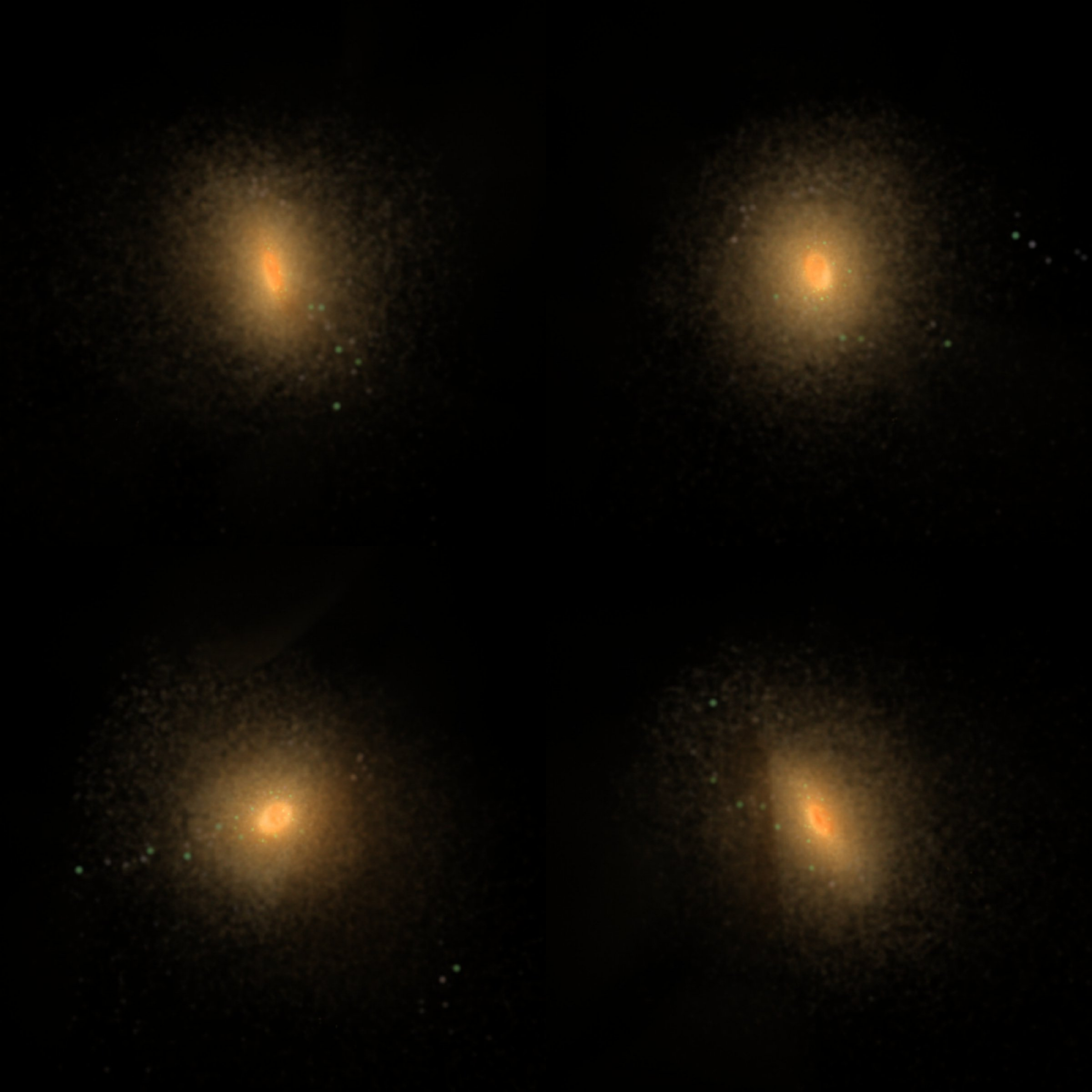}
    \end{minipage}

	\caption{
    \textbf{Barred/bar-like-structure versus unbarred galaxies in LTGs and ETGs.}
    We compare barred/bar-like-structure and unbarred Late-Type Galaxies (LTGs) and Early-Type Galaxies (ETGs).
    \textbf{Left Columns:} Three-view-projecting stellar mass surface density maps overlaid with the measured bar extent (black dashed line) for barred cases. The adjacent panels show angular velocity profiles: the orange dashed line represents the circular frequency $\Omega_{\rm circ}$ derived from the potential; grey points show the azimuthal velocity of stellar particles $v_\varphi/R$; the red solid line indicates the global pattern speed $\Omega_p$ and length measured via the method of \citet{2023MNRAS.518.2712D}; the blue curve traces the radially resolved local pattern speed following \citet{2026arXiv260305287D}, with light blue regions indicating unreliable measurement zones.
    \textbf{Right Columns:} Four orientations of synthetic RGB images (gri-bands) of the corresponding galaxies shown in the left, generated via SKIRT radiative transfer \citep{2024A&A...683A.181B}.
    \textbf{Row 1 (Barred LTG, ID 574286):} A rotation-supported disc ($D/T=0.65$) hosting a fast bar. The pattern speed (blue/red) is distinct from the material speed, characteristic of a classic density wave. Visually, it appears as a barred spiral.
    \textbf{Row 2 (Barred ETG, ID 26):} A dispersion-dominated spheroid ($D/T=0.16$) hosting a slow bar-like structure. Despite the low rotation of the stellar material (grey points well below $\Omega_{\rm circ}$), a coherent, flat pattern speed is detected (blue line). Observationally, this system lacks spiral arms and resembles an S0/E galaxy, complicating visual classification.
    \textbf{Row 3 \& 4 (Unbarred LTG/ETG):} Unbarred Late-Type/Early-Type Galaxies for comparison.
    }
	\label{fig:case_study_comparison}
\end{figure*}

Figure~\ref{fig:case_study_comparison} contrasts a canonical ``barred spiral" (ID 585282, top row) with a disputed ``bar-like ETG" (ID 26, bottom row). We visualize their mass distribution, intrinsic kinematics, and mock observational appearance based on broadband imaging from the TNG50-SKIRT Atlas \citep{2024A&A...683A.181B}.

\subsubsection{The Canonical Benchmark: ID 574286}

We select Galaxy ID 574286 (Figure~\ref{fig:case_study_comparison}, top row) as a textbook example of a barred Late-Type Galaxy (LTG). Its structural parameters—a kinematic disc-to-total ratio of $D/T=0.62$, a low S\'ersic index of $n=0.5$, and a concentration index of $R_{90}/R_{50}=2.39$—place it firmly within the disc regime.

The kinematic analysis (top-left panel) reveals the classic signature of a bar driven by orbital resonance. The grey points, representing the tangential velocity of stellar particles ($v_\varphi/R$), cluster tightly around the circular frequency curve derived from the potential (orange dashed line), confirming a rotationally supported system. Crucially, the pattern speed of the non-axisymmetric structure (blue and red lines) is distinct from the material speed. We observe a flat, coherent pattern speed $\sim 37.1\pm 0.7$ km s$^{-1}$ kpc$^{-1}$ roughly constant (using the method from \citet{2026arXiv260305287D}, with the method from \citealt{2023MNRAS.518.2712D} for crosscheck) with radius out to $\sim 2.2$~ kpc. This separation between the pattern speed ($\Omega_{p}$) and the stellar streaming motions ($v_\varphi/R$) identifies this strictly as a density wave, where stars move through the potential well of the bar.

The mock LSST observations (top-right) confirm that this dynamic structure translates to a visible bar in the optical regime, provided the viewing angle is favourable. To generate the synthetic observations, we adopted the methodology of \citet{2024A&A...683A.181B}, utilizing the \textsc{skirt} radiative transfer code to produce mock CCD images for the Legacy Survey of Space and Time (LSST). Subsequently, we constructed RGB composite images from the $g$, $r$, and $i$ bands using the mapping technique described by \citet{2004PASP..116..133L}, incorporating the considerations outlined in \citet{2019ApJ...873..111I}. For the visualization parameters, we adopted a softening parameter of $Q=8$ and a stretch value of $0.2$.

\subsubsection{The Controversy: ID 26}

In contrast, Galaxy ID 26 (second row) represents the population of odd bar-like structures characteristic of TNG50. Morphologically, it is an Early-Type Galaxy: it is dominated by random motions ($D/T=0.16$), possesses a steep light profile ($n=2.31$), and is highly concentrated ($R_{90}/R_{50}=4.37$). In standard observational classification, this system would likely be labelled a featureless elliptical or S0.

However, the kinematic decomposition (second left) reveals a robust rotating structure. Unlike the LTG case, the stellar material is largely dispersion-supported, with the average particle velocity (grey points) falling significantly below the circular velocity curve (orange line). Yet, emerging from this hot stellar bath is a coherent signal: the local pattern speed (blue line, $22 \pm 0.5$ km s$^{-1}$ kpc$^{-1}$, calculated following \citealt{2026arXiv260305287D}) exhibits a stable, flat plateau extending to $\sim 3.2$ kpc. 

This confirms that the elongation seen in the mass map is not a static triaxial ellipsoid nor a transient merger remnant, but a tumbling dynamical structure with a well-defined pattern speed. Comparisons with the LTG case reveal two key physical differences predicted in Section~\ref{sec:results}:
\begin{enumerate}
    \item \textbf{Length:} The ETG bar-like structure is physically larger, extending beyond 3 kpc compared to 2.2 kpc in the disc case.
    \item \textbf{Speed:} The ETG bar-like structure is significantly slower. While the LTG bar rotates at $\sim 40$ km s$^{-1}$ kpc$^{-1}$, the ETG bar rotates at $\sim 20$ km s$^{-1}$ kpc$^{-1}$.
\end{enumerate}

The mock images (second right) illustrate why these structures are sources of tension. In the end-on SKIRT image, the galaxy appears like S0, E or a cigar-like galaxy at different orientations. 

This case study supports the hypothesis that the ``excess" bar-like structures in TNG50 are not numerical artifacts, but rather represent a phase of secular evolution where bars persist (and grow) even as the host galaxy quenches and the disc heats up. The bar transitions from a fast density wave in a cold disc (ID 574286) to a slow, substantial triaxial tumbler in a hot spheroid (ID 26), remaining dynamically distinct but observationally camouflaged.

\subsubsection{Bar-like Structures or Simply Ellipsoidal Spheroids?} \label{sec:ellipsoidal_vs_bar}

\begin{figure*}
    \centering
    % Panel (a): Barred LTG
    \begin{minipage}{0.7\textwidth}
        %\textsf{\textbf{(a)}} Canonical Barred LTG (ID: 574286)\\
        \includegraphics[width=\textwidth]{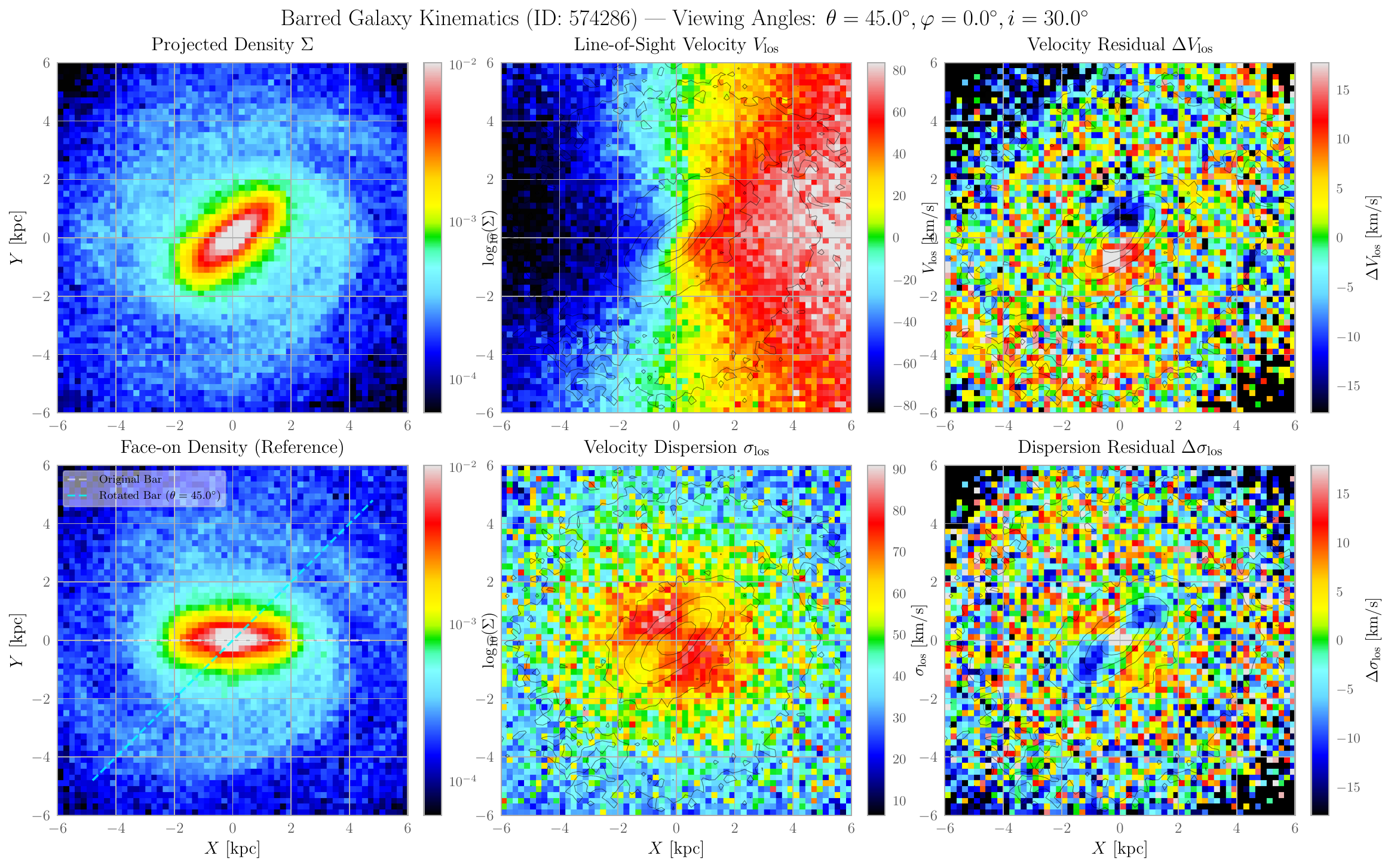}
    \end{minipage}
    
    \vspace{0.2cm} % Add some vertical spacing between the two figures
    
    % Panel (b): Bar-like ETG
    \begin{minipage}{0.7\textwidth}
        %\textsf{\textbf{(b)}} Bar-like Structure in ETG (ID: 26)\\
        \includegraphics[width=\textwidth]{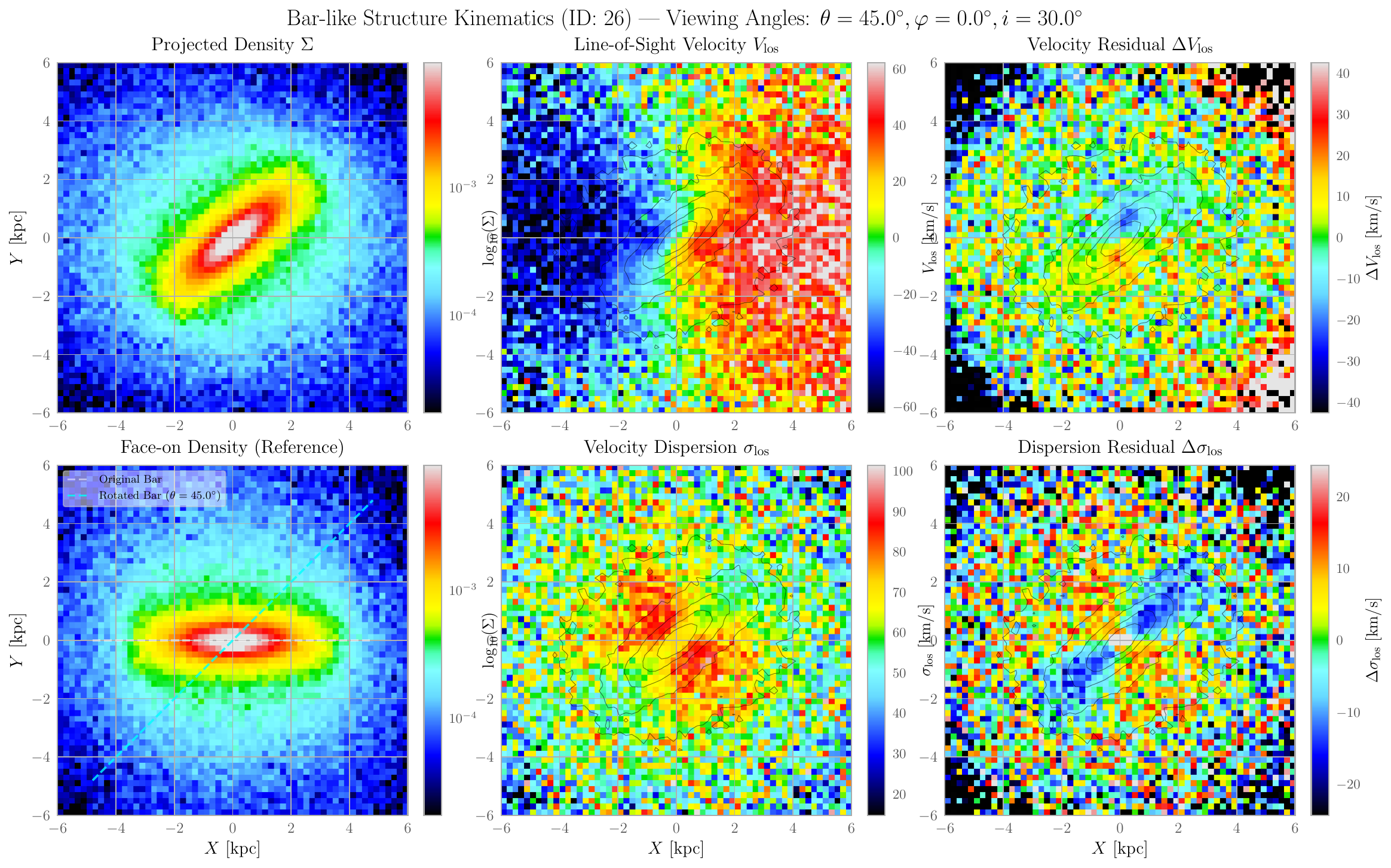}
    \end{minipage}

    \caption{
    \textbf{Spatially resolved line-of-sight (LOS) stellar kinematics and non-axisymmetric residuals.} 
    We compare the 2D kinematic structures of \textbf{(a)} a canonical barred late-type galaxy (LTG, ID 574286) and \textbf{(b)} a dispersion-dominated early-type galaxy (ETG, ID 26) hosting a secularly evolved bar-like structure. Both systems are oriented to a nearly face-on view with an inclination $i=30^{\circ}$, zero azimuthal twist ($\varphi=0^{\circ}$), and the bar's major axis rotated to a position angle of $\theta=45^{\circ}$. 
    \textit{Left columns:} Projected stellar mass surface density ($\Sigma$, top) and the perfect face-on reference density (bottom), with lines indicating the original and rotated bar axes. 
    \textit{Middle columns:} The LOS mean velocity field ($V_{\rm los}$, top) and the velocity dispersion field ($\sigma_{\rm los}$, bottom), overlaid with logarithmically spaced density contours (black lines). Both objects distinctly exhibit the characteristic $S$-shaped zero-velocity twist indicative of $x_1$ orbits responding to a barred potential. 
    \textit{Right columns:} Purely non-axisymmetric residual maps, $\Delta V_{\rm los}$ (top) and $\Delta \sigma_{\rm los}$ (bottom), generated by subtracting a 1D idealised axisymmetric reference model from the total maps. The $\Delta V_{\rm los}$ map reveals a clear kinematic dipole across the bar's minor axis. Crucially, the $\Delta \sigma_{\rm los}$ map uncovers localised kinematically cold regions at the extremities of the bar (`$\sigma$-hollows'), demonstrating that the bar-like feature in the ETG is a genuinely trapped dynamical structure rather than a purely dispersion-supported, featureless triaxial ellipsoid.
    }
    \label{fig:kinematic_maps}
\end{figure*}

A critical caveat when identifying bars in simulations, particularly within Early-Type Galaxies (ETGs) characterized by a low kinematic disc-to-total ratio ($D/T$), is the inherent limitation of the Fourier analysis method. Specifically, estimating the amplitude of the $m=2$ Fourier mode within circular annuli can yield strong artificial signals for intrinsically ellipsoidal or triaxial structures. Spheroidal systems are primarily kinematically hot and supported by velocity dispersion rather than rotation. Consequently, a static or transient triaxial ellipsoid could easily satisfy the Fourier $A_2$ criterion and be mistaken for a genuine galactic bar.

To unequivocally distinguish genuine bar-like structures from simple ellipsoidal morphologies, we rely on two robust kinematic diagnostics.

First, we utilize the mathematically rigorous framework of the local pattern speed, as detailed in our companion paper \citep{2026arXiv260305287D}. An intrinsically static or purely continuous triaxial ellipsoid lacks a coherent, solid-body tumbling motion. As showcased in Figure~\ref{fig:case_study_comparison} (blue curves), the local pattern speed profile of the bar-like structure in the ETG (ID 26) exhibits a remarkably flat and extended plateau, completely analogous to the canonical barred LTG counterpart (ID 574286). This plateau confirms that these structures are not stationary ellipsoidal features, but rather rigidly uniformly tumbling non-axisymmetric density waves. Routine inspections across our entire identified sample consistently reveal this local pattern speed plateau, confirming their nature as coherent, rotating dynamical structures.

Second, we analyze the highly detailed, spatially resolved line-of-sight (LOS) kinematic maps of these systems. Genuine bars imprint unmistakable signatures on the stellar velocity and velocity dispersion fields when observed at moderate inclinations. Figure~\ref{fig:kinematic_maps} presents the LOS velocity ($V_{\rm los}$) and velocity dispersion ($\sigma_{\rm los}$) fields, alongside their corresponding non-axisymmetric residual maps ($\Delta V_{\rm los}$ and $\Delta \sigma_{\rm los}$), for both the canonical barred LTG (ID 574286) and the controversial ETG (ID 26). To construct these mock observations, the galaxies were oriented to a nearly face-on view (inclination $i=30^\circ$, with the major axis of the structure rotated to $\theta=45^\circ$), and an idealized 1D axisymmetric kinematic model was subtracted from the data to highlight non-axisymmetric kinematic deviations.

An examination of these kinematic planes yields the following supporting evidence:
\begin{enumerate}
    \item \textbf{Kinematic $S$-shape Distortion and Velocity Dipole:} In the $V_{\rm los}$ maps, both the classic LTG bar and the ETG bar-like structure display a characteristic ``$S$-shape'' twist in the zero-velocity curve, a well-documented hallmark of stellar orbits responding to a barred potential \citep{2015MNRAS.451..936S}. When the generic axisymmetric bulk rotation is subtracted, the purely non-axisymmetric residual map ($\Delta V_{\rm los}$) exhibits a clear dipole pattern across the bar's minor axis, exactly matching observational signatures of bars \citep[see e.g.,][]{2022ApJ...939...40L}.
    
    \item \textbf{Kinematically Cold Bar Ends ($\sigma$-hollows):} Intriguingly, the dispersion residual maps ($\Delta \sigma_{\rm los}$) reveal strongly localized drops in velocity dispersion directly at the extremities of the bar structures. These kinematically cold tips is similar to the ``$\sigma$-hollow'' phenomena frequently discussed in the literature regarding double-barred galaxies \citep[e.g.,][]{2008ApJ...684L..83D, 2012MNRAS.420.1092D, 2016ApJ...828...14D}. These localized minima are generally understood to be generated by the dynamical contrast between a hotter spheroidal bulge and the relatively colder, trapped orbits of the bar. The parallel appearance of this feature in both ID 574286 and ID 26 strongly argues against the ETG structure being a homogeneously hot ellipsoid.
\end{enumerate}

We note that intrinsically strong, vertically buckling bars viewed face-on can also present a butterfly/quadrupole pattern in the mean radial velocity field \citep{2021ApJ...909..125X}. While such features represent further evidence of heavily trapped $x_1$ orbits, their manifestation is highly sensitive to the exact evolutionary phase (e.g., during active buckling) and is beyond the scope of this general census. 

Nevertheless, the ubiquitous presence of the pattern speed plateau, the $S$-shaped $V_{\rm los}$ distortions, and the localized $\sigma$-cold bar ends across our low-$D/T$ cases collectively provide definitive proof. These ``excess'' structures identified in TNG50 ETGs are indisputably genuine, secularly evolved bar-like structures, rather than false positives triggered by generic ellipsoidal shapes.

\subsubsection{Tracing the Progenitor: ID 26 at $z=0.2$} \label{sec:evolution_trace}

\begin{figure*}
	\centering
	\includegraphics[width=0.95\textwidth]{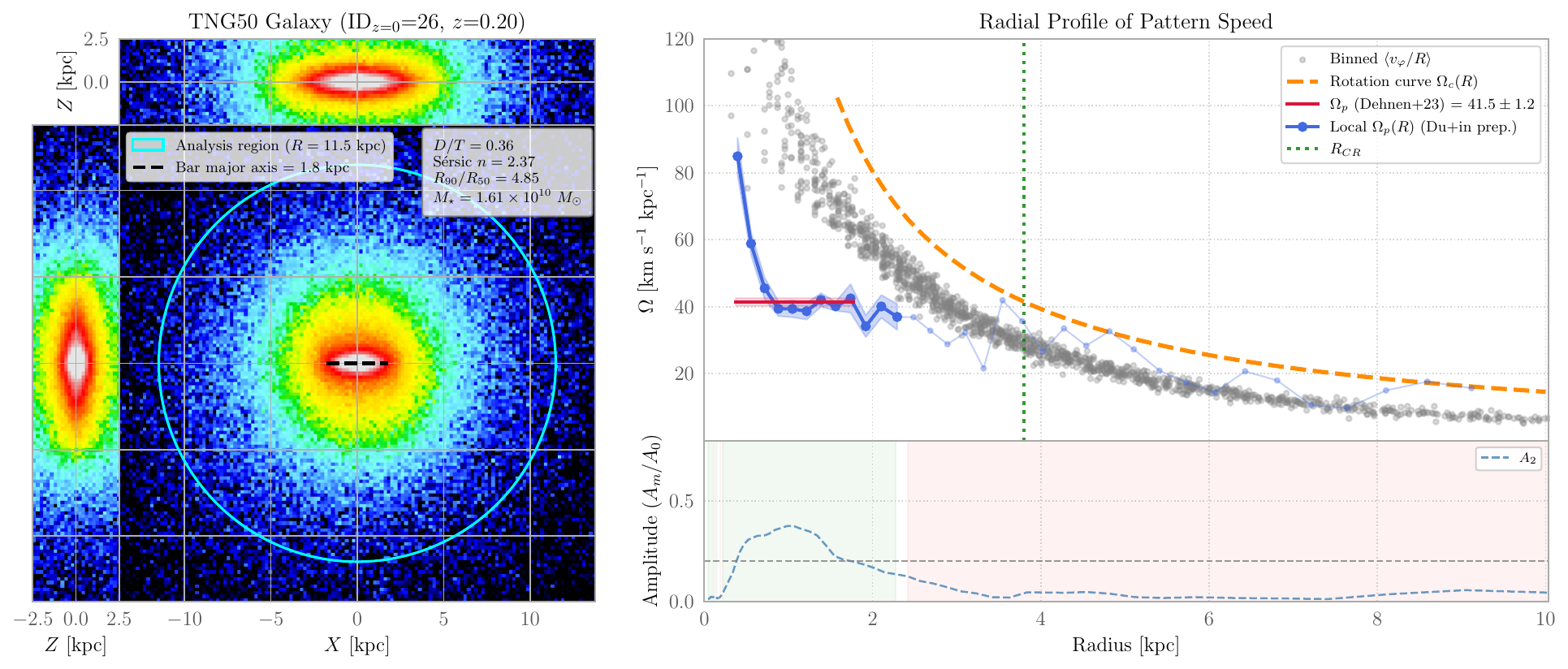}
	\caption{
    \textbf{The Progenitor of the ETG Bar: ID 26 at $z=0.2$.}
    The same kinematic and morphological analysis as Figure~\ref{fig:case_study_comparison}, but applied to the progenitor of Galaxy ID 26 at redshift $z=0.2$.
    \textbf{Left:} Three-view-projecting mass map showing a clear, short bar ($R_{\rm bar} \approx 1.8$ kpc) embedded in a disc with spiral-like features. The kinematic disc-to-total ratio is $D/T=0.36$.
    \textbf{Right (Top):} The pattern speed measurement reveals a fast-rotating structure with $\Omega_p \approx 41.5$ km s$^{-1}$ kpc$^{-1}$ (solid red line), significantly faster than its $z=0$ descendant. The grey points indicate the stellar material is more rotationally supported than at $z=0$, though significant dispersion is already present.
    \textbf{Right (Bottom):} The amplitude of the $m=2$ Fourier mode ($A_2$, blue dashed line) confirms the presence of a strong non-axisymmetric feature.
    Comparing this with the $z=0$ state (Fig.~\ref{fig:case_study_comparison}, bottom) reveals a clear evolutionary track: the ``ETG bar-like structure'' is simply a standard bar that has slowed down and lengthened over cosmic time.
    }
	\label{fig:evolution_link}
\end{figure*}

To determine the origin of these slowly rotating, elongated structures in ETGs, we trace the progenitor of Galaxy ID 26 back to $z=0.2$ ($t_{\rm lookback} \approx 2.4$ Gyr). Figure~\ref{fig:evolution_link} presents the identical analysis applied to the galaxy's progenitor state.

Strikingly, at $z=0.2$, this system exhibits the canonical properties of a barred spiral galaxy. Morphologically, the stellar distribution shows a shorter, faster bar component typical of late-type systems. Kinematically, the galaxy is significantly more rotation-supported ($D/T = 0.36$) compared to its $z=0$ state ($D/T=0.16$), although it is widely recognized that TNG50 discs tend to be kinematically hotter than real galaxies \citep[e.g.][]{2020A&A...641A..60P}.

Most crucially, the bar properties differ fundamentally from the $z=0$ epoch:
\begin{enumerate}
    \item \textbf{Pattern Speed:} The structure at $z=0.2$ is a fast rotator. The pattern speed is measured at $\Omega_p \approx 41.5 \pm 1.2$ km s$^{-1}$ kpc$^{-1}$ (red/blue lines), more than double the speed observed at $z=0$ ($\sim 15$ km s$^{-1}$ kpc$^{-1}$).
    \item \textbf{Extent:} The bar is physically shorter, with a semi-major axis of $R_{\rm bar} \approx 1.8$ kpc, compared to the extended $\sim 3.2$ kpc structure seen at $z=0$.
    \item \textbf{Disc Dominance:} The rotation curve $\Omega_{\rm circ}$ (orange dashed line) is steeper, and the stellar particles (grey dots) follow the rotation curve more closely than in the dispersion-dominated remnant at $z=0$.
\end{enumerate}

This direct evolutionary link provides unambiguous evidence that the anomalous ``ETG bar-like structure'' is not an intrinsic triaxial instability of the spheroid, but the fossilized remnant of a standard galactic bar. Over the course of $\sim 2.4$ Gyr, as the host galaxy quenched and morphologically transformed into an ETG, the bar did not dissolve. Instead, it underwent substantial secular evolution: it slowed down (losing angular momentum) and grew longer, consistent with the classical picture of resonant interaction with the dark matter halo \citep{2003MNRAS.341.1179A}. 

This suggests that TNG50 ETGs are acting as a repository for evolved bars: structures that formed in earlier, gas-richer phases and survived the quenching process to become the slow, long, camouflaged features we detect at $z=0$.

\subsubsection{Population-wide Evolutionary Trends} \label{sec:population_evolution}

\begin{figure}
	\centering
	\includegraphics[width=\columnwidth]{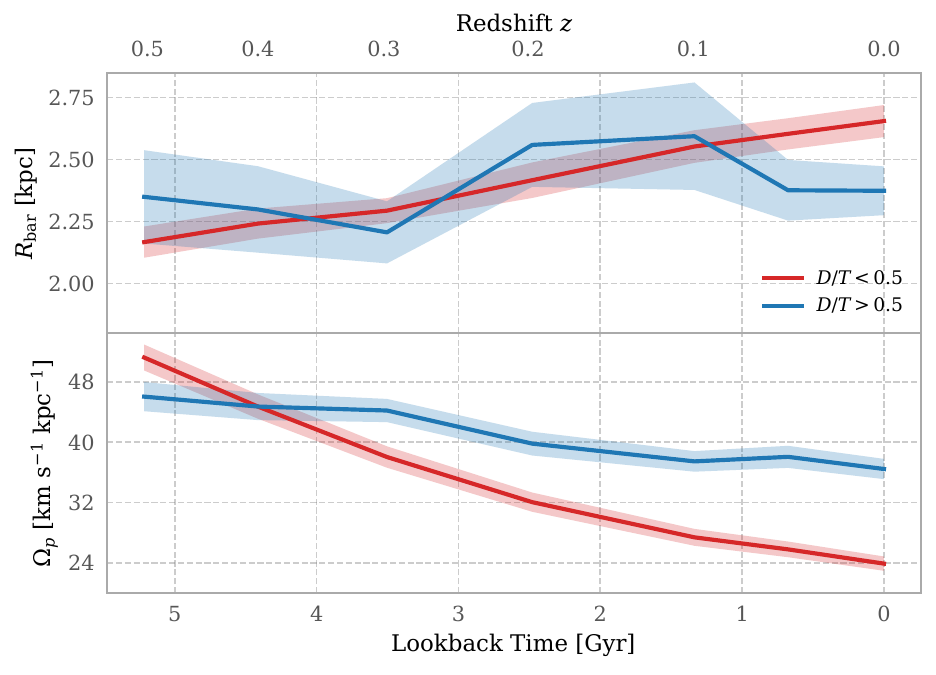}
	\caption{
    \textbf{Statistical structural and kinematic evolution of the barred population since $z \approx 0.5$.} 
    The sample of $z=0$ bar-like structures is divided into two populations based on their $z=0$ kinematic disc fraction: heavily bulge-dominated/ETG-like systems ($D/T < 0.5$, red) and disc-dominated/LTG-like systems ($D/T > 0.5$, blue). The solid lines represent the mean values for the bar semi-major axis ($R_{\rm bar}$, top panel) and the pattern speed ($\Omega_p$, bottom panel), while the shaded regions indicate the standard error of the mean. Consistent with the individual case study, the low-$D/T$ systems at $z=0$ host structures that were rotating much faster in the past, confirming their origin as classical, fast-rotating bars that have experienced significant secular braking and lengthening.
    }
	\label{fig:bar_evolution}
\end{figure}

The evolutionary pathway proposed in Section~\ref{sec:evolution_trace} for Galaxy ID 26—whereby a fast, canonical bar embedded in a disc decelerates and lengthens into a slowly tumbling triaxial structure within a quenched ETG—provides a compelling explanation for individual cases. However, to rigorously validate this mechanism and ensure it is not merely an anomaly, we must establish whether this scenario holds generally across the full simulated population.

To this end, we extend our evolutionary analysis to the complete sample of $z=0$ barred and bar-like galaxies identified in our census. By utilizing the \textsc{SubLink} merger trees, we trace the main progenitor branch of each $z=0$ barred system back to $z \approx 0.5$ (a lookback time of approximately 5 Gyr). At each available snapshot, we extract the stellar particles of the progenitor, re-align the galaxy to a face-on orientation, and employ the same Fourier-based methodology previously described to measure the instantaneous bar semi-major axis ($R_{\rm bar}$) and pattern speed ($\Omega_p$). 

To distinguish between the evolutionary tracks of typical barred spirals and the controversial bar-like structures in ETGs, we bifurcate our $z=0$ sample using a kinematic threshold of $D/T = 0.5$. This cut separates rotation-supported disc galaxies ($D/T > 0.5$) from dispersion-dominated, spheroidal systems ($D/T < 0.5$), following the criteria frequently adopted in the literature to isolate galactic discs in cosmological simulations \citep[e.g.,][]{2022MNRAS.512.5339R}.

Figure~\ref{fig:bar_evolution} presents the population-averaged evolutionary tracks for both length (top panel) and pattern speed (bottom panel). Moving forward in cosmic time (from left to right), two prominent features immediately emerge:

First, both populations exhibit a ubiquitous, overarching trend of secular evolution characterized by concurrent bar lengthening and deceleration. Over the last $\sim 5$ Gyr, the mean bar pattern speeds systematically drop, while the physical extents of the bars slowly grow. This general behaviour of continuous angular momentum loss and bar growth within the TNG50 simulation is entirely consistent with recent, independent structural analyses \citep[e.g.,][]{2024A&A...691A.122H, 2024A&A...692A.159S}.

Second, and most crucially for identifying the origin of ETG bars, the $D/T < 0.5$ population exhibits an extreme phase of deceleration. At higher redshifts ($z \sim 0.5$), the progenitors of today's slowly tumbling, dispersion-dominated bar-like structures possessed remarkably fast pattern speeds—averaging $\Omega_p \gtrsim 50$ km s$^{-1}$ kpc$^{-1}$—which were, in fact, initially higher than those of typical disc galaxy progenitors. As these host galaxies evolved over cosmic time (a period during which many transition morphologically from discs to ETGs due to processes such as mergers and AGN feedback), their bars did not dissolve. Instead, they endured severe dynamical braking, losing immense amounts of angular momentum, likely to the dark matter halo and the growing stellar spheroid, eventually settling into the slow ($\Omega_p \sim 25$ km s$^{-1}$ kpc$^{-1}$), elongated structures we observe at $z=0$.

This statistical confirmation robustly solidifies the hypothesis drawn from our case studies. The ``excess" of bar-like features identified in $TNG50$'s low-rotation regime are not numerical noise, nor are they generic, featureless ellipsoids born natively out of hot stellar orbits. They are the fossilized, secularly evolved remnants of historically fast, canonical bars that have profoundly transformed in tandem with their host galaxies.

\section{Discussion} \label{sec:discussion}

The results presented in this work reveal a fundamental feature of the TNG50 simulation: the ubiquity of bar-like structures in systems kinematically and morphologically classified as Early-Type Galaxies. By adopting a continuous, morphology-agnostic view, we have shown that bar-like structures in TNG50 are not restricted to cold discs but persist—and indeed dominate—in the dispersion-dominated regime. Here, we discuss the physical drivers of this phenomenon, the tension it creates with standard observational paradigms, and whether it represents a numerical artefact or a genuine prediction of galaxy evolution.

\subsection{Prolate Rotators or Not?} 
\label{sec:prolate}

Before the systematic discussion, we note a complicating factor in interpreting these elongated ETGs is the observational existence of prolate rotators. IFS surveys such as CALIFA \citep{2017A&A...606A..62T} and MUSE \citep{2018MNRAS.477.5327K} have revealed that prolate rotation (angular momentum aligned with the distinct, long axis) is more common than previously thought, particularly in most massive galaxies like BCGs, reaching fractions of up to 50\%. Crucially, however, we find a distinct kinematic mismatch in TNG50. Using the angular momentum determination method described in Section~\ref{sec:frame}, we find no instances of genuine prolate rotation (rotation about the major axis, \citealt{2017A&A...606A..62T}) in our sample; the elongated TNG50 galaxies discussed here are exclusively oblate rotators (rotating about the minor axis), consistent with bar-like kinematics rather than prolate spheroids. This suggests that the TNG ``bar-like" ETGs are physically distinct from the massive prolate rotators observed in the Universe.

\subsection{Formation: Low Spin as the Driver} 
\label{sec:discussion_formation}

At first glance, the prevalence of bar-like structures in ETGs ($D/T < 0.2$) appears counter-intuitive given the canonical association of bars with rotationally supported discs. However, our findings align robustly with theoretical expectations governing stability and angular momentum.

It has long been established that low angular momentum facilitates bar instabilities. Theoretically, a decrease in rotational support relative to random motion does not necessarily stabilize a system against non-axisymmetric modes; rather, high spin is the stabilizing factor that prevents bar formation \citep{1969ApJ...155..393P, 1982MNRAS.199.1069E, 2014ApJ...783L..18L}. It is worth noting that while theoretically sound, classical definitions like the ELN82 instability criterion have been shown to face significant scaling challenges when tested directly against multi-wavelength observations of local disc galaxies \citep{2023MNRAS.518.1002R}. Simulations have repeatedly demonstrated that in high-spin environments, bar formation is suppressed or results only in weak geometric distortions \citep{2003MNRAS.341.1179A, 2018MNRAS.476.1331C}. While massive spheroids can facilitate bar growth via angular momentum exchange, potentially leading to stronger bars in bulge-dominated systems \citep{2002MNRAS.330...35A, 2003MNRAS.341.1179A,2012MNRAS.425L..10S}.

This inverse correlation is clearly visible in specific observational regimes. For instance, Low Surface Brightness (LSB) galaxies, which are typically gas-rich and possess high specific angular momentum ($\lambda_R$), exhibit a remarkably low bar fraction \citep{2017ApJ...847...37C, 2025MNRAS.539.2262C}. Conversely, observational studies focusing on the secular evolution of bars suggest that strong bars are preferentially hosted by systems with lower spin parameters \citep{2013ApJ...775...19C, 2014ApJ...785..137S}, a trend recently reinforced by \citet{2026MNRAS.546ag015A}, who report a clear anti-correlation between bar fraction and $\lambda_{R}$. Observations consistently show that bars are prevalent in massive, red, bulge-dominated galaxies exhibiting low specific star formation rates and gas-poor conditions \citep{2012ApJ...750..141L, 2012MNRAS.424.2180M, 2012MNRAS.423.1485S, 2016A&A...595A..67C, 2013ApJ...779..162C, 2017ApJ...835...80C}.

In this context, TNG50 is behaving consistently with dynamical theory. The ``excess" bar-like structures in ETGs reside in gas-poor, lower-spin systems where the stabilizing influence of a cold, high-angular-momentum gas disc is absent. The tension, therefore, is not between the simulation and physics, but between the simulation and the morphological definitions typically applied to observational data.

\subsection{Evolution: The Secular Braking of Bars}
\label{sec:bar_braking}

The transformation of Galaxy ID 26 from a fast-barred progenitor at $z=0.2$ to a slow, elongated rotator at $z=0$ (Fig.~\ref{fig:evolution_link}) serves as a textbook example of secular evolution driven by dynamical friction.

The secular evolution of galactic bars is fundamentally governed by the exchange of angular momentum with the host galaxy, particularly through resonant interaction with the dark matter halo. Foundational theoretical works established that this interaction transfers angular momentum from the bar to the halo, inevitably driving bar deceleration \citep{1972MNRAS.157....1L, 1984MNRAS.209..729T, 1985MNRAS.213..451W}. This mechanism was rigorously substantiated by N-body simulations, which indicated that the braking process is inextricably coupled with bar growth: as bars lose angular momentum, they become longer and stronger \citep{2003MNRAS.341.1179A}. 

Our findings in TNG50 mirror this dynamical prediction. The bars in ETGs are universally slower and longer than their counterparts in LTGs, consistent with them being dynamically ``older'' structures that have experienced prolonged friction against the halo. Crucially, once established, these non-axisymmetric structures exhibit remarkable durability, hard to be destroyed \citep{2004ApJ...604..614S}. 

Consequently, the abundance of bar-like structures in TNG50 ETGs may simply reflect the endpoint of secular evolution where bar destruction mechanisms are inefficient. If TNG50 galaxies quench and settle early—effectively creating the ``low spin'' conditions required for formation (Section~\ref{sec:discussion_formation})—the resulting bars will persist indefinitely. In this view, the simulation presents a ``future universe'' scenario: as the cosmetic gas supply of the real universe dwindles and secular evolution proceeds, the population of long, slow bars in quiescent galaxies is expected to rise, eventually matching the demographics seen in TNG50.

\subsection{Agreement on the ``Red Branch''}
\label{sec:red_branch}

Before attributing statistical discrepancies solely to systematic biases, it is crucial to establish where the simulation physically aligns with observations. While the global bar fraction in TNG50 ETGs appears high, the trends governing the presence of these bar-like structures show a remarkable consistency with the ``quiescent branch'' of observational censuses.

As detailed in Section~\ref{sec:compare_with_observation}, the probability of hosting a bar in TNG50 correlates positively with stellar mass (peaking at $\sim 10^{10.8} \rm M_{\odot}$), redder colours, and higher concentration indices ($C = R_{90}/R_{50}$). They mirror specific observational realities. For instance, the unimodal peak in the mass dependence aligns with the high-mass behavior reported in Galaxy Zoo Hubble \citep[][Fig.5]{2014MNRAS.438.2882M} and the S$^4$G survey \citep[][Fig.19]{2016A&A...587A.160D}.

More importantly, some surveys reveal a bimodal reality in the relation of bar fraction against Hubble type, color, and concentration (\citealp[][Fig.4]{2010ApJ...714L.260N}; \citealp[][Fig.7]{2015ApJS..217...32B}; \citealp[][Fig.10]{2016A&A...587A.160D}; \citealp[][Fig.8]{2022MNRAS.512.2222V}; \citealp[][Fig.~6]{2025MNRAS.542..151M}). They all show that, while the bar fraction drops for intermediate types, it rises again or remains significant for the earliest types (S0--Sb, larger $g-r$ or concentration). This suggests that the TNG50 bar-like structures are physically analogous to the structures found in the massive, red, high-concentration end of reliability-tested observational catalogues.

Nevertheless, while the simulation successfully recovers the high bar incidence characteristic of this ``red branch,'' the global demographic topology presents a more complex picture. TNG50 does not fully reproduce the distinct \textit{bimodality}—specifically the profound ``dip'' at intermediate types—seen in the aforementioned censuses. These discrepancies indicate that, while TNG50 robustly captures the physics of secular evolution in massive, hot systems, the precise demographic balance—particularly regarding the ``valley'' between the blue and red populations—remains sensitive to both the simulation's quenching timescale and the observational classification pipeline employed.

\subsection{The Green Valley ``Dip'': A Missing Evolutionary Link?}
\label{sec:green_valley}

As highlighted in Section~\ref{sec:red_branch}, while TNG50 successfully reproduces the high bar incidence in the ``red branch,'' it predicts a strictly monotonic increase in $f_{\rm bar}$ with $g-r$ colour (Figure~\ref{fig:obs_comparison_grid}b). This monotonic trend stands in contrast to the bimodal distributions reported in several observational censuses, which identify a profound ``dip'' in bar fraction among intermediate-type galaxies. Physically, this intermediate regime corresponds to the ``green valley'' (GV)---the transitional phase between the star-forming blue cloud and the quiescent red sequence.

Recent observational studies have explicitly quantified this structural deficit in the transition zone. For instance, \citet{2021JCAP...06..045D} found that only $\sim 6\%$ of green valley galaxies host a bar, suggesting that the majority of GV galaxies curtail their star formation via smooth, mass-driven internal processes rather than violent bar- or merger-driven mechanisms. Similarly, morphological analyses of the GV reveal a prevalence of fading disc features, such as rings and loosening spiral arms, rather than prominent bars \citep{2022MNRAS.517.4575S}.

When bars \textit{are} present in the green valley, they appear to dictate a highly specific quenching pathway. \citet{2019A&A...630A..88N} demonstrated that barred GV galaxies exhibit significantly longer (slower) quenching timescales compared to their unbarred counterparts, indicating that the bar's presence is a morphological signature for slow, secular quenching rather than rapid, violent truncation. Furthermore, \citet{2025A&A...696A.118R} showed that bars in this transitional phase drive ``inside-out'' quenching; by funneling gas to the galactic centre, bars exhaust the inner fuel supply, turning the inner regions (up to the bar length) red and passive while the extended disc may still retain some star formation. This theoretical picture is beautifully corroborated by detailed multi-wavelength observations of individual systems \citep[e.g.][]{2019A&A...621L...4G}.

The juxtaposition of these observational facts with our TNG50 results reveals a critical nuance in the simulation's evolutionary pathways. In TNG50, we observe a continuous survival of bars: fast bars in blue discs ($z \sim 0.2$) simply decelerate and lengthen as the galaxy quenches into a red ETG ($z=0$, see Section~\ref{sec:evolution_trace}). The lack of a ``GV dip'' in TNG50 implies that simulated bars might survive the morphological and quenching transition too efficiently. 

In the real Universe, the low bar fraction in the GV suggests two possibilities: either (a) pre-existing fast bars in blue discs are frequently destroyed or dissolved during the transition to the red sequence (perhaps via minor mergers or specific feedback mechanisms not fully captured in TNG50), requiring new bars to reform later in the red sequence; or (b) the transition through the green valley is so rapid for strongly barred galaxies that they are under-represented in observational snapshots. Given that TNG50 bars act as ``indestructible fossils'' (Section~\ref{sec:bar_braking}), the simulation heavily favors a continuous, non-destructive secular evolution. This over-preservation of bars through the green valley phase likely contributes directly to the over-production of the slow, bar-like structures we observe in the $z=0$ ETG population.

\subsection{The Divergence of Observational Pipelines}
\label{sec:observation_bias}

Given the demographic consistency described above, the stark quantitative disagreement in the bar fraction might stems from a divergence in classification paradigms.

The majority of large-scale observational censuses \citep[e.g.][]{2011MNRAS.411.2026M} implicitly adopt the logic of the traditional Hubble Tuning Fork. In this framework, the presence of a disc is a prerequisite for a bar; galaxies are often pre-sorted into Discs or Spheroids, and only the former are visually inspected for bars, which totally rules out the possibility of the existence of bar-like structures in ETGs. Consequently, an ETG with a bar-like structure is strictly an oxymoron in many automated pipelines—it is classified as an Elliptical due to its bulge dominance, and the bar is lost to the designation. This might explain why MaNGA DR17 \citep[e.g.,][Fig.18]{2025MNRAS.544.1056V} and Galaxy Zoo DECaLS \citep{2022MNRAS.509.3966W} report a strictly unimodal distribution peaking at Late-Types (Sab) with a sharp decline towards S0s.

However, when one adopts the Comprehensive VRHS Classification system \citep{1979ApJ...227..714K, 2011A&A...532A..74B}, which treats the Hubble Type ($T$) and the Bar Family ($F$) as orthogonal parameters, the picture reconciles with our findings. Using this framework on mid-IR imaging, \citet{2015ApJS..217...32B} and \citet{2016A&A...587A.160D} report the same bimodal distribution hinted at by our concentration trends. Crucially, they identify a non-negligible fraction of barred galaxies at $T \le 0$ (S0s and Ellipticals). Similarly, the ${\rm ATLAS}^{\rm 3D}$ project \citep{2011MNRAS.414.2923K}, with similar philosophy, revealed that bars and/or rings are present in 30\% of ETGs. This motivates \citet{2019MNRAS.487.4995G} to raise the concept of a galaxy classification framework that better recognizes bars in ETGs. What is more, \citet{2010ApJ...714L.260N, 2015ApJS..217...32B, 2016A&A...587A.160D, 2022MNRAS.512.2222V} all reported significant bimodal distributions in bar fractions, with two sub-samples exhibiting diametrically opposed trends.

The bar fraction bias introduced by observational effects has also been discussed \citep[e.g.,][]{2025PASA...42..166I, 2026MNRAS.546ag263G}.

\subsection{A New ``Bar Problem'' or ETG Over-production?}
\label{sec:new_problem}

Despite the theoretical plausibility, we cannot dismiss the possibility that TNG50 produces these structures too efficiently. The historical ``missing bar problem" in cosmological simulations \citep{2017MNRAS.469.1054A, 2019MNRAS.483.2721P} was solved in IllustrisTNG through enhanced feedback mechanisms. It is conceivable that the pendulum has swung too far (or other imperfect physics effects), creating a ``New Bar Problem" where bars are over-stabilized in quenched systems. 

This tension is objectively quantifiable in galaxy kinematics. As noted by \citet{2020A&A...641A..60P}, TNG50 ETGs frequently populate the observational ``forbidden region'' on the $\lambda_{R} - \varepsilon$ diagram (high ellipticity but low rotational support). They found that this discrepancy is driven precisely by the population of prolate or bar-like structures in ETGs identified in this work; removing them reconciles the simulation with the ATLAS$^{\rm 3D}$ domain. 

This could also be symptomatic of ``over-quenching" reported in other contexts by \citet{2020MNRAS.499.4239S, 2021MNRAS.502.3685A},  where the simulated quiescent fraction exceeds empirical observations by a factor of 2-5. This discrepancy is attributed to the subgrid physics governing AGN feedback, specifically the adopted threshold halo mass for black hole seeding and the initial black hole seed mass. Furthermore, foundational scaling relations critical for bar formation, such as the stellar-to-halo mass fraction, are known to be systematically underpredicted in simulations like IllustrisTNG and EAGLE \citep{2020MNRAS.499.5656R}. The same numerical recipes often result in simulated discs that are overly stable against gravitational instability \citep[with little variance in the Toomre parameter;][]{2020MNRAS.499.5656R}. This aligns with other noted tensions in the simulation, such as the ``Red Disc" excess \citep{2019MNRAS.483.4140R} and subtle chemical mismatches in quenched populations \citep{2021MNRAS.501.4359Z}. Moreover, the importance of both subgrid physics and environmental history to bar formation in IllustrisTNG is highlighted by \citet{2025A&A...698A..20R} and \citet{2024A&A...684A.179R}.

We note the frequently reported issue of bars rotating too slowly \citep{2017MNRAS.469.1054A, 2019MNRAS.483.2721P} or extending too short \citep[\textit{f} panel of our Fig.\ref{fig:obs_comparison_grid};][]{2022ApJ...940...61F, 2020ApJ...904..170Z} in cosmological simulations—once considered a tension between observations and $\Lambda$CDM—was alleviated in the Auriga simulations through higher resolution and refined feedback models \citep{2021A&A...650L..16F}, demonstrating that the discrepancy arose from overly efficient stellar and AGN feedback mechanisms in EAGLE or Illustris.

\section{Conclusions} \label{sec:conclusions}

We have presented a comprehensive analysis of bar-like structures in the TNG50 simulation at $z=0$, aiming to resolve the nature of the elongated morphologies frequently identified in its Early-Type Galaxy population. By moving beyond disc-only pre-selection, we recover a continuous demographic of non-axisymmetric structures. Our main conclusions are as follows:

\begin{enumerate}
    \item \textbf{Ubiquity in ETGs:} TNG50 predicts a high prevalence of bar-like structures ($f_{\rm bar} \sim 75-80\%$) in massive, red, old, gas poor, dispersion-dominated galaxies ($D/T < 0.2$). These are not distinct populations from the classical barred spirals but form a continuum in parameter space (see Fig.\ref{fig:stacked_profiles}~\&~\ref{fig:obs_comparison_grid}).
    
    \item \textbf{Dynamical Reality:} These structures are not static triaxial ellipsoids or merger remnants, neither prolate galaxies inside which stars rotate around the long axis. Kinematic decomposition reveals they are genuine rotating patterns with coherent pattern speeds. By tracing progenitors, we confirm they are evolved remnants of standard bars that have undergone significant secular braking and lengthening ($R_{\rm bar} \gtrsim 3$ kpc, $\Omega_{\rm p} \sim 10-20$ km s$^{-1}$ kpc$^{-1}$, see Fig.\ref{fig:case_study_comparison}~\&~\ref{fig:evolution_link}).
    
    \item \textbf{Consistency with Kinematic Theory:} The rise of bar fraction in TNG50 ETGs is consistent with the angular momentum evolution of galaxies. We find a strong anti-correlation between rotational support and bar presence, supporting theoretical predictions and observations (see section~\ref{sec:discussion_formation}) that low-spin systems are highly susceptible to bar instabilities, whereas high-spin systems suppress them.

    \item \textbf{Agreement with Observational Trends:} The demographic trends in TNG50 mirror the ``red branch'' of observations. The bar fraction rises with stellar mass (peaking at $\sim 10^{10.8} \rm M_{\odot}$), deeper red colours, and higher concentration indices. This matches the specific bimodal behaviour reported in recent surveys \citep[e.g.,][]{2016A&A...587A.160D,2022MNRAS.512.2222V,2025MNRAS.542..151M} when early-type populations are isolated  (see section~\ref{sec:discussion_formation}~\&~\ref{sec:red_branch}). 

    \item \textbf{Non-negligible tensions:} Despite the qualitative match on the ``red branch'', significant tensions remain. TNG50 predicts a monotonic increase in bar fraction with colour, failing to reproduce the bimodal distribution and the distinct ``dip'' observed in intermediate-type, green valley galaxies \citep[e.g.,][]{2021JCAP...06..045D}. Furthermore, these bar-like structures populate the observational ``forbidden region'' of the $\lambda_{R} - \varepsilon$ diagram \citep{2020A&A...641A..60P}, and simulated bars are systematically shorter than observed \citep[see our Fig.\ref{fig:obs_comparison_grid}f;][]{2020ApJ...904..170Z, 2022ApJ...940...61F}. This indicates that TNG50 may be over-preserving bars during the quenching transition, leading to an over-production of these structures in ETGs, possibly linked to the AGN feedback subgrid physics (see section~\ref{sec:green_valley} \& \ref{sec:new_problem}).

    \item \textbf{The Fossil Record of Discs:} The ubiquity of these structures suggests that bar destruction is inefficient in the TNG50 physics model. Once formed in the star-forming phase, bars survive the quenching transition. The excess of ETG bar-like structures in TNG50 implies that simulated galaxies may form bars too efficiently in ``warm'' discs due to rapid angular momentum loss, and these bars subsequently serve as indestructible fossils of the galaxy's disc past (see section~\ref{sec:evolution_trace}~\&~\ref{sec:bar_braking}).
    
\end{enumerate}

We conclude that the ``excess'' bar-like structures in TNG50 likely reflect a convolution of two factors: an imperfect baryonic physics model that may over-producing bars or ETGs, and a significant population of real-world secular structures that remain misclassified in standard morphological censuses (see section~\ref{sec:observation_bias}). Future work must bridge the semantic gap between the kinematic definitions of structures in simulations and the photometrically driven taxonomy of observations, ensuring that the ``E'' and ``S0'' classifications do not artificially obscure the dynamical richness of the early-type galaxies. We are currently addressing the potential observational biases in a companion study.

\section*{Acknowledgements}

We thank Youjun Lu and Shude Mao for their expert guidance and valuable discussions that greatly enhanced this work. 
%We thank the anonymous referee for their constructive comments. 
We also extend our gratitude to E. Athanassoula for her support during the early stages of this project. This work is supported by the National Key Research and Development Program of China (No. 2023YFA1607904), the National Astronomical Observatories of the Chinese Academy of Sciences (No. E4ZR0510), the Beijing Municipal Natural Science Foundation (No. 1242032), the Youth Innovation Promotion Association of the Chinese Academy of Sciences (No. 2022056), and the China Manned Space Program with grant no. CMS-CSST-2025-A07. The IllustrisTNG simulations were undertaken with compute time awarded by the Gauss Centre for Supercomputing (GCS) under GCS Large-Scale Projects GCS-ILLU and GCS-DWAR on the GCS share of the supercomputer Hazel Hen at the High Performance Computing Center Stuttgart (HLRS), as well as on the machines of the Max Planck Computing and Data Facility (MPCDF) in Garching, Germany. We are grateful for the use of the TNG project's online JupyterLab service and high-performance computing resources, which greatly facilitated the data analysis presented in this work.

%%%%%%%%%%%%%%%%%%%%%%%%%%%%%%%%%%%%%%%%%%%%%%%%%%
\section*{Data Availability}

The data underlying this article will be shared on reasonable request to the corresponding author.



\bibliographystyle{mnras}
\bibliography{example} % if your bibtex file is called example.bib

%\begin{thebibliography}{99}
%\end{thebibliography}

%%%%%%%%%%%%%%%%%%%%%%%%%%%%%%%%%%%%%%%%%%%%%%%%%%

%%%%%%%%%%%%%%%%% APPENDICES %%%%%%%%%%%%%%%%%%%%%

%\appendix


%If you want to present additional material which would interrupt the flow of the main paper, it can be placed in an Appendix which appears after the list of references.

%%%%%%%%%%%%%%%%%%%%%%%%%%%%%%%%%%%%%%%%%%%%%%%%%%

% Don't change these lines
\bsp	% typesetting comment
\label{lastpage}
\end{document}